\documentclass[10pt, journal]{IEEEtran}
\usepackage{cite}
\usepackage{amsmath,amssymb,amsfonts}
\usepackage{newtxtext}
\usepackage{newtxmath} 
\usepackage{algorithmic}
\usepackage{graphicx}
\usepackage{textcomp}
\renewcommand{\thetable}{\Roman{table}}
\usepackage{url}
\makeatletter
\AtBeginDocument{\DeclareMathVersion{bold}
\SetSymbolFont{operators}{bold}{T1}{times}{b}{n}
\SetMathAlphabet{\mathrm}{bold}{T1}{times}{b}{n}
\SetMathAlphabet{\mathit}{bold}{T1}{times}{b}{it}
\SetMathAlphabet{\mathbf}{bold}{T1}{times}{b}{n}
\SetMathAlphabet{\mathtt}{bold}{OT1}{pcr}{b}{n}
\SetSymbolFont{symbols}{bold}{OMS}{cmsy}{b}{n}
\renewcommand\boldmath{\@nomath\boldmath\mathversion{bold}}}
\makeatother

\def\BibTeX{{\rm B\kern-.05em{\sc i\kern-.025em b}\kern-.08em
    T\kern-.1667em\lower.7ex\hbox{E}\kern-.125emX}}

\title{Neural Equalisers for Highly Compressed Faster-than-Nyquist Signalling: Design, Performance, Complexity and Robustness }
\author{\uppercase{Shubham Paul }\IEEEauthorrefmark{1}, \IEEEmembership{Student Member, IEEE},
\uppercase{Sheetal Kalyani }\IEEEauthorrefmark{1}, \IEEEmembership{Member, IEEE},
\uppercase{Nambi Seshadri }\IEEEauthorrefmark{2} \IEEEauthorrefmark{1}, \IEEEmembership{Fellow, IEEE}, and \uppercase{R David Koilpillai}\IEEEauthorrefmark{1},
\IEEEmembership{Senior Member, IEEE}
\IEEEauthorblockA{\IEEEauthorrefmark{1}Indian Institute of Technology Madras}
\IEEEauthorblockA{\IEEEauthorrefmark{2}Jacobs School of Engineering, UC San Diego}
}
\begin{document}

\maketitle

\begin{abstract}
Faster-than-Nyquist (FTN) signalling has emerged as a compelling technique for enhancing spectral efficiency in bandwidth-constrained communication systems. By intentionally introducing controlled intersymbol interference (ISI), FTN allows transmission at rates exceeding the traditional Nyquist limit, unlocking new potential in high-speed data communication. However, its practical deployment remains challenged by the need for low-complexity detection strategies that can cope with the induced ISI while maintaining low latency and robust performance. We propose deep learning receivers that are resilient to non-idealities.

In this paper, we present a deep learning-based framework for FTN signalling that addresses these challenges through several novel contributions. First, we propose a sliding window detection method that leverages temporal context while preserving computational efficiency. Second, we demonstrate the viability of FTN systems with very low packing factors, showing that reliable performance can be achieved even under aggressive spectral compression (up to 75\%). Our architecture is optimised for low latency and low complexity, making it suitable for real-time applications and scalable deployment. In addition, we assess the robustness of our models across varying channel conditions and noise profiles, providing insights into their generalisability and resilience.

\end{abstract}


\section{INTRODUCTION} \label{sec: Introduction}
\IEEEPARstart{S}{ince} the 1970s, the technique of Faster-than-Nyquist (FTN) signalling\cite{Tufts_0, Liveris, Mazo} has offered the potential for transmitting information at rates higher than the conventional Nyquist rate for digital communications. 
With symbol intervals shorter than those specified by the Nyquist criterion, FTN provides higher transmission rates by introducing intentional Inter-Symbol Interference (ISI) at the transmitter. 
However, this requires more complex detection techniques at the receiver. 

The problem of designing receivers for FTN systems has drawn significant interest from the research community since its inception. \cite{Tufts_0, Mazo, Mazo_FTN_min_dis, Hajela_FTN_min_dis, FDE_FTN, SVD_FTN, Liveris,   Prlja_turbo,  Prlja_Reduced, Anderson_turbo, FTN_Anderson,Jana_Precoding, FTN_overview, Eigen_FTN_IM}. 
It was demonstrated in \cite{FTN_Anderson}, that the gains in spectral efficiency are more pronounced for small values of the FTN packing factors ($\tau$) such as 0.25.
But, these cases with higher rate gains introduced significantly higher ISI. 
The decoding of such signals needed the use of trellis structures, which were computationally expensive.
Detectors for FTN signals with small $\tau$ were developed in \cite{Prlja_turbo, Prlja_Reduced, Anderson_turbo, FTN_Anderson}, using a trellis-based BCJR~\cite{BCJR_OG} algorithm.
The works also introduced an efficient algorithm for determining the Minimum Euclidean Distance (MED) which was used to obtain lower bounds on BER performance. 
The bounds are presented in \cite{Prlja_turbo}. 
However, the reduced-complexity BCJR receivers are still computationally complex and have a degradation in BER with respect to the lower bounds. 
As a result, subsequent works were mostly limited to mild interference scenarios arising from large to moderate $\tau$ (>0.5).
With the advent of Deep Learning (DL) techniques, we ask the following: Can we design low complexity DNN Rx for small $\tau$? We then provide an affirmative answer to the same and also demonstrate that these proposed Rxs are quite resilient to non-idealities.

\footnote{ Preliminary results of this work were presented at the 2023 IEEE International Conference on Advanced Networks and Telecommunications Systems (ANTS) in \cite{Self_1}.}

More recently, there has been a growing interest in leveraging ML algorithms for signal processing and physical layer (PHY) problems. 
DL-based methods to design (coded) modulation, and multiple access for Simultaneous Wireless Information-Power Transmission (SWIPT) networks were explored in \cite{ML_for_Wireless}. 
Communications system design was interpreted an end-to-end reconstruction task that seeks to jointly optimise transmitter and receiver components, as an autoencoder was presented in \cite{DL_for_PHY}. 
The effort was further expanded to facilitate the joint optimisation of bit-wise mutual information (BMI) in \cite{ML_Constellations}. 
In \cite{BPTA}, a practical method, based on stochastic perturbation techniques, was developed to train an end-to-end communication system without relying on explicit channel models. 
A generative supervised deep neural network for 1-bit quantised OFDM signals was presented in \cite{Onebit_OFDM}. 

Deep Learning (DL) equalisers learn the complex ISI patterns to mitigate the impact of the ISI. 
A DL-based receiver design to equalise and detect the data symbols in molecular communication was given in \cite{Molecular_DL}.
ML-based symbol detection for finite-memory causal channels based on the Viterbi algorithm was developed in \cite{ViterbiNet}.
ViterbiNet integrated DNN-based sequence classifiers to perform CSI-dependent computations in the Viterbi algorithm, leveraging the Markovian structure of finite-memory causal channels. 
Deep learning based detection using a Sliding Bidirectional Recurrent Neural Network (SBRNN) was proposed in \cite{SBRNN_Transactions}.
The SBRNN was demonstrated to be computationally efficient and could perform detection under various channel conditions without knowledge of the underlying channel model.

\begin{table}
    \centering
    \begin{tabular}{|c|c|c|c|}\hline
         Paper&  High & Deep  &Sliding \\
         & Compression& Learning&window\\\hline
 Sugiura (2013) \cite{FDE_FTN}& No& No&No\\\hline
 Nie (2016) \cite{Interference_cancellation}& No& No&No\\\hline
 Bedeer (2017)\cite{Bedeer}& No& No&No\\\hline
         Wen (2018, 2022)\cite{Message_passing, FTN_precoding_preequalization}& No & No & No\\\hline
         Song (2020) \cite{FTN_SIC}& No& Yes& No\\\hline
         Sina (2022) \cite{FTN_LSD}& No& Yes& No\\\hline
         Liu (2021)  \cite{FTN_Sum_Product}&  No&  Yes&No\\\hline
         Yang (2024) \cite{FTNOWC_MIMO_IC}&  No&  Yes&No\\\hline
         Baek (2024) \cite{FTN_Quasi}&  No&  Yes&No\\\hline
 Anderson (2010, 2013) \cite{Anderson_turbo, FTN_Anderson}& Yes& No&No\\\hline 
 Prlja (2008, 2010) \cite{Prlja_turbo, Prlja_Reduced}& Yes& No&No\\\hline
 Current paper& Yes& Yes&Yes\\ \hline
    \end{tabular}
    \caption{Comparison of some relevant works}
    \label{tab: Comparison of relevant works}
\end{table}

We compare our approach to some recent works on FTN in Table~\ref{tab: Comparison of relevant works}. An FTN receiver using a Generalised Message Passing (GMP) algorithm with a factor graph was proposed in  \cite{Message_passing}, and Successive Interference Cancellation (SIC) based receivers were studied in \cite{Bedeer} and \cite{Interference_cancellation}. 
However, such methods were prohibitively complex even for moderate compressions. 
To manage the detection complexity of FTN signalling, low-complexity frequency domain equalisers (FDEs) were explored in \cite{FDE_FTN, FTN_CP, FTN_precoding_preequalization}. 
These works demonstrated reliable performance at low complexity for low compression but suffered for moderate and high compressions. 
A hybrid approach for FTN signal detection, combining DL with SIC was suggested in \cite{FTN_SIC}. 
A joint signal detection and decoding scheme using DL was also proposed. 
In \cite{FTN_LSD}, a DL-based List Sphere Decoding (DL-LSD) algorithm was introduced, reducing computational complexity compared to the original LSD algorithm. 
In \cite{FTN_Sum_Product}, a detection algorithm based on a modified factor graph, which concatenates a neural network function node, was proposed. 
Variable-Packing-Ratio (VPR)-based transmissions for high spectrum efficiency and security were proposed in \cite{FTN_Variable_Packing_Ratio}. 
Deep learning solutions for FTN in the context of Optical Wireless Communication (OWC) systems were explored in \cite{FTNOWC_MIMO, FTNOWC_MIMO2,FTNOWC_MIMO_IC,FTNOWC_PEq}. 
With \cite{FTNOWC_MIMO,FTNOWC_MIMO2,FTNOWC_MIMO_IC} focussed on the development of deep learning-based FTN detectors with MIMO and \cite{FTNOWC_PEq} aimed at developing a deep learning-based pre-equalisation scheme.
ML-based signal detection for FTN in quasi-static channels was explored in \cite{FTN_Quasi}. 
The  LSTM-based FTN detector proposed in \cite{FTN_Quasi} demonstrated the ability to detect transmitted symbols without estimating channel coefficients and SNR. 
However, the focus of the references mentioned earlier was mild interference scenarios arising from large to moderate $\tau$ (i.e, low and moderate compressions, respectively).

Table~\ref{tab: Comparison of relevant works} contrasts the current work against the background literature. 
The delay and latency exhibited by trellis-based receivers can be circumvented by using RNNs. 
Such solutions had performances comparable to the trellis-decoders but at a drastically lower complexity. 
What also sets us apart is the sliding window detection mechanism with the RNNs.
This work continues on the same signal and receiver model. 

The major contributions of this paper are:
\begin{enumerate}
    \item \textbf{Small Packing Factors} This paper provides an in-depth demonstration of FTN packing factors $\tau \leq 0.5$ that enable the system to achieve data transmission rates that exceed twice those possible via conventional orthogonal signalling methods. 
    However, the equalisation schemes discussed in~\cite{FDE_FTN, FTN_CP, FTN_precoding_preequalization} start to show a significant performance degradation for the packing factors $\tau < 0.7$ as shown in Fig. 7 of reference \cite{Evolution}.
Thus, these receivers are inadequate for small FTN packing factor scenarios. 
    The existing literature indicates that Trellis-based structures like those in~\cite{Anderson_turbo, Prlja_Reduced} and~\cite{Prlja_turbo} present the only viable solutions currently available for ISI mitigation in these conditions.
This study aims to introduce and analyse alternative receiver architectures designed to tackle ISI more effectively.

    \item \textbf{Deep Neural Network Receivers}: In this work, various DNN receivers are developed.

    \begin{enumerate}
        \item \textbf{Sliding Window Neural Network Receiver}:We propose a sliding window receiver developed in \cite{Self_1}\footnote{This is our conference paper} is extended to multiple NN architectures. 
        The NN models function as non-linear equalisers.
        We discuss the architecture in detail and contrast the performance of the various NN architectures.
        After the training phase, these models have low computational complexity when in operation and at the same time, maintain good BER performance. 
        The advantages of this system, including its adaptability and efficiency, are elaborated upon in Section IV. 
    
        \item \textbf{Detection with Feedback: } Models without feedback underperformed in certain scenarios, as elaborated in the upcoming sections. 
        Thus, models that incorporate feedback mechanisms are also proposed in this paper.
        Specifically, Bi-RNNs are used to implement feedback, improving performance metrics at the cost of a manageable reduction in complexity.
        This feedback approach improves detection accuracy as it offers contextual information from past and future symbol sequences.
        
        \item \textbf{Attention Mechanisms:} We also approach the equalisation problem by modelling it as a sequence-to-sequence learning task, where transformers are employed as an alternative to RNNs. 
        We present preliminary results on the use of transformer architecture to detect FTN signals.
        
        Attention mechanisms allow the model to focus on relevant parts of the input sequence, enhancing its performance and adaptability in various contexts.
    \end{enumerate}

    \item \textbf{Testing under Non-Ideal Conditions:} To assess the robustness of our NN models, comprehensive testing under various non-ideal conditions, including sampling offsets, and mismatch in training Signal-to-Noise Ratio (SNR) is conducted. 
    The observations illustrate that the proposed receivers possess the capacity for generalisation, allowing them to maintain BER performance even in the face of non-idealities. 
    This resilience is crucial for real-world deployments.
Existing studies have not addressed the performance of the proposed receivers under real-world conditions.

    \item \textbf{Addressing Generalisation Concerns}: One of the predominant challenges associated with Deep Learning models is a potential overfitting problem, which undermines their ability to generalise to scenarios that were not represented in the training dataset. 
    Our study demonstrates that the proposed models effectively mitigate these concerns, retaining strong generalisation capabilities even when trained on a limited sample space.

\end{enumerate}

The paper is structured as follows: 
In Section~\ref{sec: System-Model-and-Problem-Formulation}, the system model is introduced and the relevant modifications are discussed.
In Sec~\ref{sec: The-NN-based-Sliding-window-detection}, we propose various neural architectures.
The architecture of the sliding NN, its training and operation is discussed in Sec~\ref{sec: selecting_NN_Arch}. 
In Section~\ref{sec: Results}, we present the performance of the equalisers designed. 
Finally, the complexity of the proposed systems is studied in Section~\ref{sec: Complexity-Analysis}.

\section{SYSTEM MODEL}
\label{sec: System-Model-and-Problem-Formulation}
\subsection{FTN Signalling Model} \label{subsec: FTN-Signalling-Problem}

\begin{figure}
\centerline{\includegraphics[width = 0.9\linewidth]{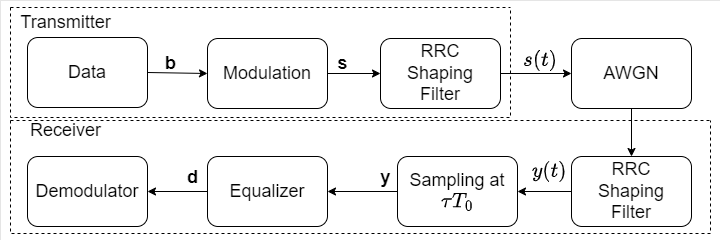}}
\caption{Generic FTN System Model.}
\label{Fig: Generic FTN System Model}
\end{figure}	
We intend to transmit real-valued symbols \(s[n]\) using a Root Raised Cosine (RRC) pulse \(q(t)\) and a symbol interval \(T_0\). The transmitted symbol is given by:
\begin{equation} 
\label{system-equation}
s(t) = \sum_{n} {s[n] q(t -  n T_0)}.
\end{equation}

In the case of an Additive White Gaussian Noise (AWGN) channel, the received signal is \(y(t)\in \mathbb{R}\) at the output of the matched filter. 
If $y(t)$ is sampled at the \(k^{th}\) intervals, then the corresponding vector \( y[k]=y(kT_0)\) obtained is given by:
\begin{align}
    y[k] &= \sum_n s[n] g((k-n) T_0 )+\eta (kT_0 ),\\
    y[k] &=s[k] g(0)+\sum _{n\neq k} s[n] g((k-n) T_0 )+\eta (kT_0), \label{sampled-equation}
\end{align}
where $g(t) = q(t)*q^*(-t)$ is a Raised Cosine (RC) pulse.

The first term in \eqref{sampled-equation} corresponds to the desired symbol. 
The second term corresponds to the ISI which is ideally zero. 
This is the Nyquist's condition for zero ISI. 
In relation to spectral requirements, it can be summarized as the following observation: 
If a channel is band-limited to a bandwidth of 2\textit{W}, the condition for zero ISI is satisfied as long as \(T_0 \geq \frac{1}{2W}\).
FTN schemes are essentially signalling schemes where the symbol interval \(T<\frac{1}{2W}\). 
The relation between the Nyquist interval and the FTN interval is given by $T=\tau T_0$ where \(0<\tau\leq 1 \).
FTN can achieve higher rates than the conventional Nyquist scheme at the cost of increased ISI. 
As $\tau$ is reduced, the ISI increases.
The vector \(y[k] = y(kT) = y(k\tau T_0)\) is obtained via sampling.
The samples, $y[k]$ may be written as a block of N symbols \(\) as 

\begin{align}
\textbf{y}=&[y_0, y_1,...,y_{N-1]}] \\
    \textbf{y}=& \textbf{G}\textbf{s}+\boldsymbol{\eta} .
\end{align}

Matrix $\textbf{G} \in \mathbb{R}^{N \times N} $  captures the impact of ISI due to FTN and $\eta (kT)$ captures noise samples. $\textbf{G}$ is given by: 

\begin{align}
\label{Matrix-equation} \textbf{G} =
\begin{bmatrix} 
g(0) &~g(-T) &~ \cdots &~ g(-(N-1)T) \\ 
g(T) &~g(0) &~ \cdots &~ g(-(N-2)T) \\ 
g(2T) &~g(T) &~ \cdots &~ g(-(N-3)T) \\ 
\vdots &~ \vdots &~ \ddots &~ \vdots \\ 
g((N-1)T) &~ g((N-2)T) &~ \cdots &~ g(0) 
\end{bmatrix},
\end{align}
and the correlation of the noise is $\mathbb{E} [ \boldsymbol{\eta \eta^H} ] = N_0 \textbf{G}$\cite{Anderson_turbo}. 

It is worth noting that $\textbf{G}  = \mathbf{I_N} $  for \(\tau =1\), i.e., Nyquist signalling. 
For $\tau \leq \frac{1}{1+\beta}$, where $\beta$ is the roll-off factor for the RRC pulse, the system encounters spectral nulls~\cite{Jana_Precoding, Prlja_turbo}. 
As such, the matrix $\textbf{G}$ is ill-conditioned or even singular, making an inversion difficult and/or impossible.

A discrete-time whitening filter is necessary at the receiver to effectively process coloured noise samples after the matched filter stage for FTN signals. 
However, devising a causal and stable discrete-time whitening filter is challenging, especially when dealing with small values of $\tau$. 
Alternative methodologies are required to overcome this challenge. 
One such option, as implemented in this study, encompasses the adoption of an equivalent FTN signalling model. 
This model leverages a set of orthonormal basis functions to achieve noise whitening after the matched filter operation.
As a result, the effective FTN channel is modified.
Details of the model are discussed in the following subsection. 
Notably, this specific model was introduced in \cite{Prlja_turbo}.

\subsection{The Modified FTN Channel } \label{subsec: The-Modified-FTN-Channel}
\begin{figure}
\centerline{\includegraphics [width = 0.9\linewidth]{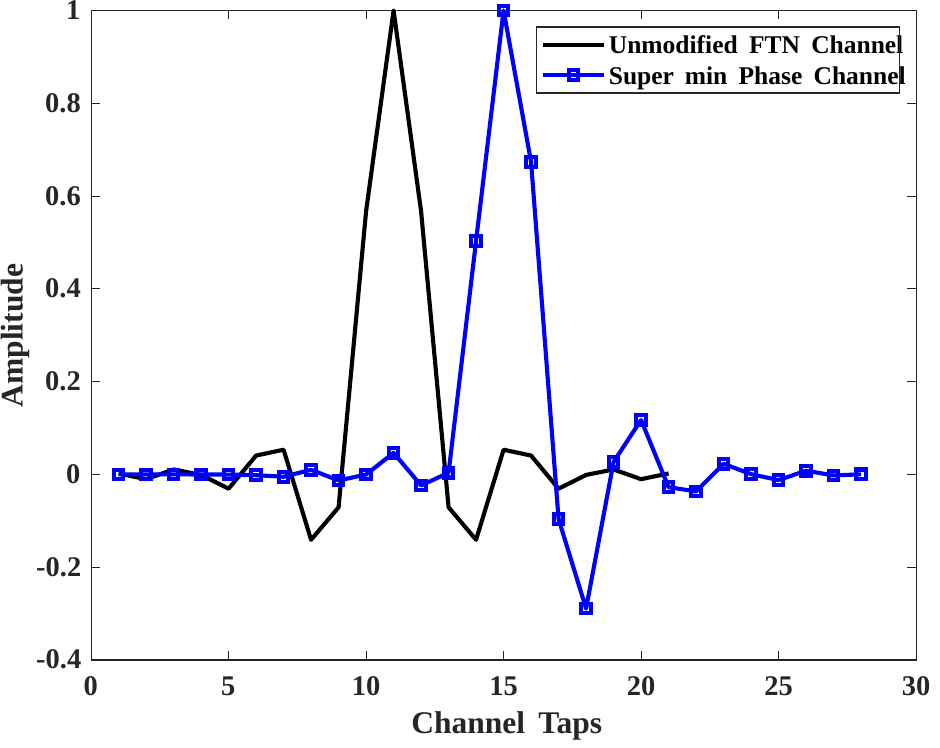}}
\caption{Comparison of FTN Channels: Unmodified and Super Minimum Phase}
\label{Fig: Comparison of FTN Channels}
\end{figure}

\begin{figure}
\centerline{\includegraphics [width = 0.9\linewidth]{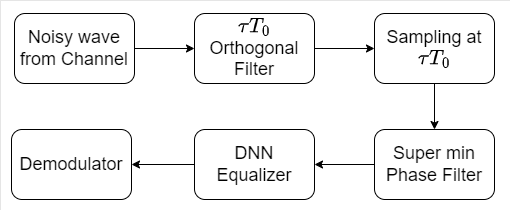}}
\caption{Modified System Diagram}
\label{Fig: Modified-System-Diagram}
\end{figure}

As shown in~\eqref{Matrix-equation}, the FTN channel can be viewed as a linear ISI channel with the ISI depending on $\tau$ and $\beta$. 
For $ \tau = 0.50$ and $\beta = 0.30$, the ISI lasts $18$ symbol intervals. 
Hence maximum likelihood sequence estimation (MLSE) is complex and requires a trellis of $ 2^{18} $ states. 
A truncated Viterbi algorithm or reduced complexity trellis search techniques benefit significantly if the channel model has a minimum phase. 
In\cite{Prlja_thesis}, the concept of a super-minimum phase channel is introduced. 
While a true minimum phase system ensures minimum group delay, a super minimum phase ensures rapid energy growth in model taps. 
The latter may produce a low-energy precursor of length $ L_P$. 
This is accounted for by introducing an offset of $ L_P$ symbols. 
However, the effective channel length is shortened.
Fig.~\ref{Fig: Comparison of FTN Channels} shows how the impulse response is altered by the super minimum phase system for  $ \tau = 0.50 $ and $\beta = 0.3$. 
Note that the super minimum phase response produces a low-energy precursor of length 13. 
This essentially means running the decoders ahead by 13 symbols.
Fig.~\ref{Fig: Modified-System-Diagram} shows the resultant modified system diagram. 
The shorter effective channel implies that the MLSE can be carried out with a reduced trellis.
Furthermore, the DFE needs fewer taps for the forward and feedback filters. 
In our simulations, we use the reduced channel. 
The reduced channel allows us to reduce the size of the input window for our models. 
This allows our models to be more computationally efficient. 
Refer~\cite{Prlja_thesis} for more details on the super-minimum phase channel.
Other works in Table~\ref{tab: Comparison of relevant works} usually use a whitening matrix or equalise coloured noise samples.

\subsection{Problem Definition} \label{subsec: Problem-Definition}
In Subsection~\ref{subsec: FTN-Signalling-Problem}, we noted that the matrix $\textbf{G}$ in \eqref{Matrix-equation}  is ill-conditioned for $\tau \leq \frac{1}{1+\beta}$. 
As such, an inverse is not feasible. 
As indicated in Section~\ref{sec: Introduction}, interesting cases of FTN signalling occur when $\tau \leq \frac{1}{1+\beta}$ and demand study. 
In light of this, we present our solution to detect FTN signals for the instances where $ \tau<\frac{1}{1+\beta}$ using the modified FTN channel discussed in \ref{subsec: The-Modified-FTN-Channel}
We explore the sequential detection of symbols as an approach to detecting FTN signals.

\section{THE NN-BASED SLIDING WINDOW DETECTION} \label{sec: The-NN-based-Sliding-window-detection}
In this section, we discuss how Deep Neural Networks (DNNs) can equalise and detect FTN signals. 
We highlight the benefits and drawbacks of various DNN techniques.
We then present and discuss the proposed sliding-window detector in detail. 

\subsection{Variation of DNN Detectors} \label{subsec: Variation-of-DNN-Detectors}
\textbf{Memoryless DNN Detectors: }The input to this framework consists of an observed symbol, denoted as \(y[k]\), while the output is the corresponding symbol label, represented as \(x[k]\). 
These detectors detect symbols individually, performing well when the channel has no memory. 
However, they struggle with the ISI caused by adjacent symbols in FTN signalling, making them unsuitable.

\textbf{Block DNN Detectors:} Block detectors handle blocks of symbols, effectively capturing relationships within them.
The input to this framework consists of an observed block of symbols, denoted as \textbf{y}, while the output gives the corresponding block of symbol labels, represented as \textbf{x}. 
The complexity of the architecture depends on the block size $L$. 
While they perform better in ISI scenarios, they cannot address IBI from overlapping blocks, compromising detection. 
Techniques like zero-padding can mitigate this but reduce efficiency and increase overhead. 
Additionally, larger block sizes raise detection complexity.

\textbf{Sliding Window DNN Detectors:}  To address IBI, sliding window detectors incorporate a sliding mechanism, allowing blocks to overlap. 
The architecture processes sequences more efficiently than traditional block detection.
Similar to the standard block detector, the input is an observed block of $L$ symbols, denoted as \textbf{y}, while the output \textbf{x} corresponds to the associated block of $l$ symbol labels. 
With a block size of $L$, as the window progresses by $l$ symbols, there exist $L-l$ overlapping symbols between successive blocks. 
Furthermore, the sliding mechanism allows predictions for subsequent outputs, waiting only for a fraction of the symbol block at the detector.
This reduces the latency of the detection.
Consequently, this functionality considerably reduces latency within the system. 
The next subsection details our proposed detector.
\begin{figure}
\centerline{\includegraphics [width = 0.9\linewidth]{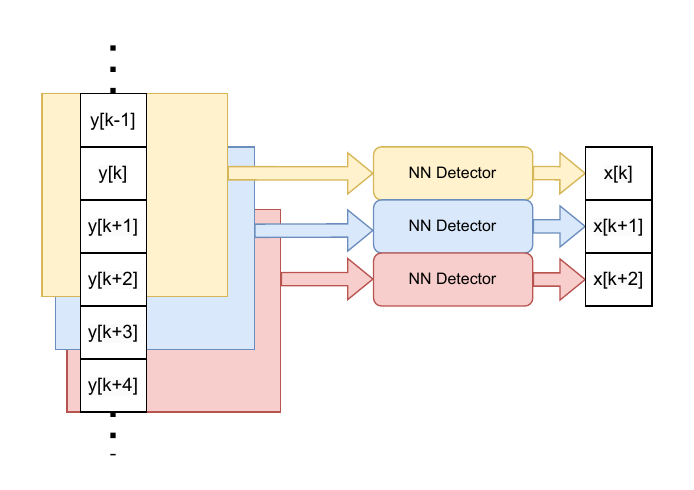}}
\caption{The operation of the Sliding NN detector}
\label{Fig: Sliding NN detector}
\end{figure}
\subsection{The Modified Sliding Window Detector: }\label{The-Sliding-Window}
In the previous subsection, we outlined the advantages and drawbacks of block detectors and sliding window detectors.
We observe the inherent simplicity of the memoryless detector and the remarkable detection capabilities of the sliding window detector in identifying symbols within a channel that displays memory characteristics. 
Our objective has been to develop a network that combines the detection proficiency of the sliding window detector with the speed and simplicity characteristic of memoryless detectors. 
We accomplish this by implementing a sliding window detector that incrementally shifts by one symbol. 
This results in a symbol-by-symbol detection process similar to that of memoryless detectors, while preserving the sliding window's capability to effectively capture ISI and IBI.

We proceed assuming that the initial symbols are known. 
As each subsequent symbol becomes available, the neural network undergoes a shift by incorporating the newly observed symbol.  Fig.~\ref{Fig: Sliding NN detector} shows the operation of the sliding window NN detector. 
The NN's role is to predict the output for individual input symbols, which effectively is a classification task.
This neural decoder can be an FCNN, an unidirectional RNN, a Bidirectional RNN or a Transformer network.
We shall explore the various architectures in the upcoming sections.
We can express the detection problem for the $k^{th}$ transmit symbol ($ \hat{x}[k] $) as follows:
\begin{equation}
\hat{x}[k] = \arg \max_{x[k]} P(x[k]|Y_k),
\label{eq: detection-equation}
\end{equation}

where,
\begin{multline}
    Y_k = \{ y[k-l_1], y[k-l_1+1], ... , y[k], \\..., y[k+l_2-1],y[k+l_2]\} 
    \label{window}
\end{multline}
    
$ Y_k $ is the $k^{th}$ observation window, $ {x}[k] \in \{-1, +1\}$  and $ \hat{x}[k] $ gives the estimate of the $k^{th}$ transmit symbol. 
The window size L is given by:
$ L = l_1 + l_2 + 1 $, where $l_1 $ denotes the number of past symbols and $l_2$ denotes the number of future symbols contributing to the ISI. 
In Fig.~\ref{Fig: Sliding NN detector}  we consider $ l_1 =1 $ and $ l_2 = 2$ as an example. 

\subsection{The Detection Problem} \label{Detection-as-a-classification-problem} 

The DNN attempts to solve the detection problem given in \eqref{eq: detection-equation}.
To perform the detection, it minimises a cross-entropy loss. 
Let the actual and estimated PMFs be given by $p(.)$ and $\hat{p}(.)$ , respectively. 
The cross-entropy loss is given by: 
 \begin{equation}
      H(p,\hat{p}) = H(p) + D_{KL}(p||\hat{p})   .
\label{eq: CE_loss}
 \end{equation}
The minimisation of cross-entropy loss plays a critical role in minimising the Kullback-Leibler divergence between the true probability mass function $(p(.))$ and the one inferred by the neural network $(\hat{p}(.))$. 

For binary models that classify observations into $0$ and $1$, the sigmoid activation function $g(z) = \frac{1}{(1 + e^{-z})}$, where $z$ is a function of the input $Y_k$, is used to model the probability of the output $x[k]$ being $1$. 
The complement of the probability obtained from the model gives the probability of $x[k]$ being $0$. 
A simple comparison of the two yields the final result.

It is worth noting that this concept can be extended to encompass higher-order modulation schemes with some modifications. 
A categorical cross-entropy loss will be required to replace the binary cross-entropy in this model. 
Additionally, utilising a one-hot representation for the output and subsequently employing a softmax operation in the final layer will be necessary to determine the output.

\section{Building and Training our Models}
\label{sec: selecting_NN_Arch}
In the previous section, we discussed the NN-based sliding window detector but the architecture of the NN model was not discussed.
In this section, we shall consider the various architectures that may be used to design the NN detector in Fig.~\ref{Fig: Sliding NN detector}.

\begin{figure}
\centerline{\includegraphics [width = 0.9\linewidth]{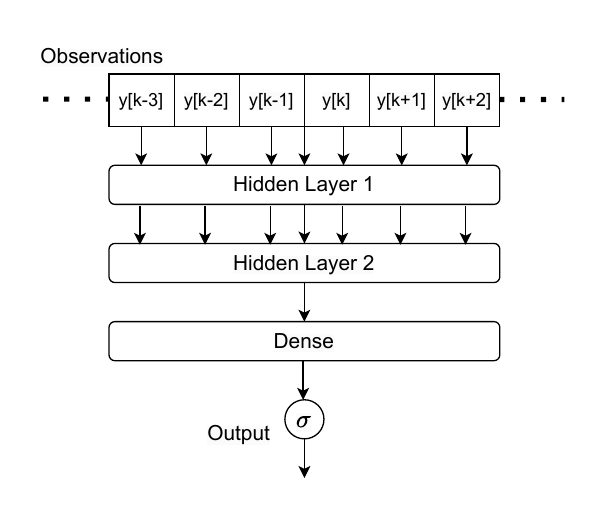}}
\caption{FTN NN Structure}
\label{fig: FTN_NN_Structure}
\end{figure}

The works presented in \cite{ViterbiNet, SBRNN_Transactions, SBRNN} indicate that a relatively shallow architecture is all that is needed for symbol detection tasks in classical  wireless communication.
This is unlike the deep networks used in image processing \cite{NIPS2012_c399862d, Image_recognition} or speech processing applications \cite{DL_speech_Acoustic}.
This could potentially be because the structure in the input signal is less complex for symbol detection problems than those encountered in images, text, or speech.
Hence, we restrict ourselves to the space of shallow networks. Shallow networks also require much less compute and memory for inference time which is desirable.
Specifically, we maintain a structure as shown in Fig.~\ref{fig: FTN_NN_Structure}.
That is, we have an input layer, 2 hidden layers and an output layer with a single node.
The specific cases of the architecture are discussed as follows.
\subsection{Fully Connected NN-based Structures}\label{Feed-Forward-NN-based-Structures}
This is the most basic NN architecture that can be employed for detection.
It uses several fully connected NN layers followed by a final softmax layer \cite{lecun2015deep,heaton2018ian}. 
This is called Fully Connected Neural Network (FCNN). 
Such networks are simple, fast and are good universal function mappers. 
Our neural network architecture, implemented using TensorFlow, comprises an Input Layer of dimension $L$ (= window size), the two hidden layers in Fig~\ref{fig: FTN_NN_Structure} are of type Dense, each featuring a size of 64, and an Output Layer consisting of a single neuron within a Dense layer activated by a sigmoid function.

\subsection{RNN-based Structures} \label{RNN-based-Structures}
It's worth noting that FCNNs  are limited in their ability to account for the temporal and sequential implications of a signal. 
This is where Recurrent Neural Networks (RNNs) excel. 
RNNs are a popular choice for sequence estimation in different problems such as neural machine translation \cite{bahdanau2014neural}, speech recognition \cite{hinton2012deep} and/or prediction tasks. 
With their ability to manage ISI \cite{SBRNN, SBRNN_Transactions}, they ensure highly accurate sequence detection. 

\textbf{Type I: Unidirectional Models}

A significant advantage of RNNs is their adaptability - they process incoming data streams in real-time, leveraging the RNN's state to integrate insights from prior observations. 
Unlike the previous subsection, the two hidden layers are now made of RNN layers, each featuring an unfolding size of 64, but the output layer remains the same.
LSTM cells yield two outputs per cell: the cell state and the hidden state. 
On the other hand, Simple RNN (SRNN) cells and Gated Recurrent Unit (GRU) cells only produce the cell state. 
By default, the RNN layer provides solely the final hidden state output. 
This output encapsulates an abstract representation of the input sequence. 
We intend to use the initial RNN layer as a sequence-to-sequence modelling component, to stack multiple RNN layers. 
Consequently, we set the \textit{return-sequences} parameter to \textit{True}, which yields the states after each time step as a sequence. 
This sequence is then directed into another RNN layer. The second RNN layer isn't stacked and furnishes solely the final state. 
The RNN layers are followed by a Dense layer with one output and a sigmoid activation.
This is used to return the probabilities. Table~\ref{tab: NN layer details} details the size and activation of the layers of the different networks. 
The value of $N$ varies in our model according to the FTN packing ratio ($\tau$). 

\textbf{Type II: Bidirectional Models}

RNNs, though promising in detecting incoming data streams at the receiver, cannot consider future symbols. 
One way to address this limitation is by using a Bi-RNN. 
In our paper, we showcase the utilisation of multiple RNN cell variants. 
The received signal sequence is forwarded into one RNN cell and backwards into another RNN cell, allowing the two outputs to be passed to additional bidirectional layers. 
This Bi-RNN architecture ensures that future signal observations are considered when estimating a symbol. 
The hidden layers in Fig.~\ref{fig: FTN_NN_Structure} are Bi-LSTM layers. 
Table~\ref{tab: NN layer details} details the size and activation of the layers of the Bi-RNN networks. 
Similar to the unidirectional case, $N$ varies according to $\tau$. That is, the smaller the $\tau$, the larger the $N$.

\begin{table*}[]
\centering
\caption{ The NN layer details}
\begin{tabular}{| c|c|l|l|c|l|l|c|c|c|c  |} \hline  

Layer& \multicolumn{7}{|c|}{Type}&Input size &Output size &Activation\\\hline  
 &  FCNN & \multicolumn{3}{|c|}{RNN Uni}& \multicolumn{3}{|c|}{RNN Bi}& & &\\ \hline  
 
  Input layer & Input    & Input    &Input    & Input    & Input    &Input    &Input   & $L$& $L$& linear \\ \hline  

 Layer 1& Dense  & SRNN&GRU& LSTM & Bi-SRNN&Bi-GRU&Bi-LSTM& $L$& $N$&ReLU\\ \hline  

 Layer 2& Dense & SRNN&GRU& LSTM & Bi-SRNN&Bi-GRU&Bi-LSTM&  $N$& 1 &ReLU\\ \hline  

 Output Layer& Dense   & Dense&Dense   & Dense   & Dense   &Dense   &Dense  & 1 & 1 & Sigmoid\\ \hline

\end{tabular}
\label{tab: NN layer details}
\end{table*}

\subsection{Attention Models: TransformerNet}

\begin{figure}
\centerline{\includegraphics [width = 0.9\linewidth]{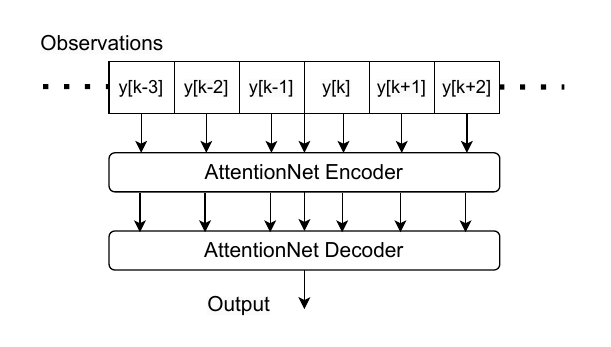}}
\caption{Simplified \textit{FTN-TransformerNet} Structure}
\label{fig: Simplified FTNTransformerNet Structure}
\end{figure}

\begin{figure}
\centerline{\includegraphics [width = 0.9\linewidth]{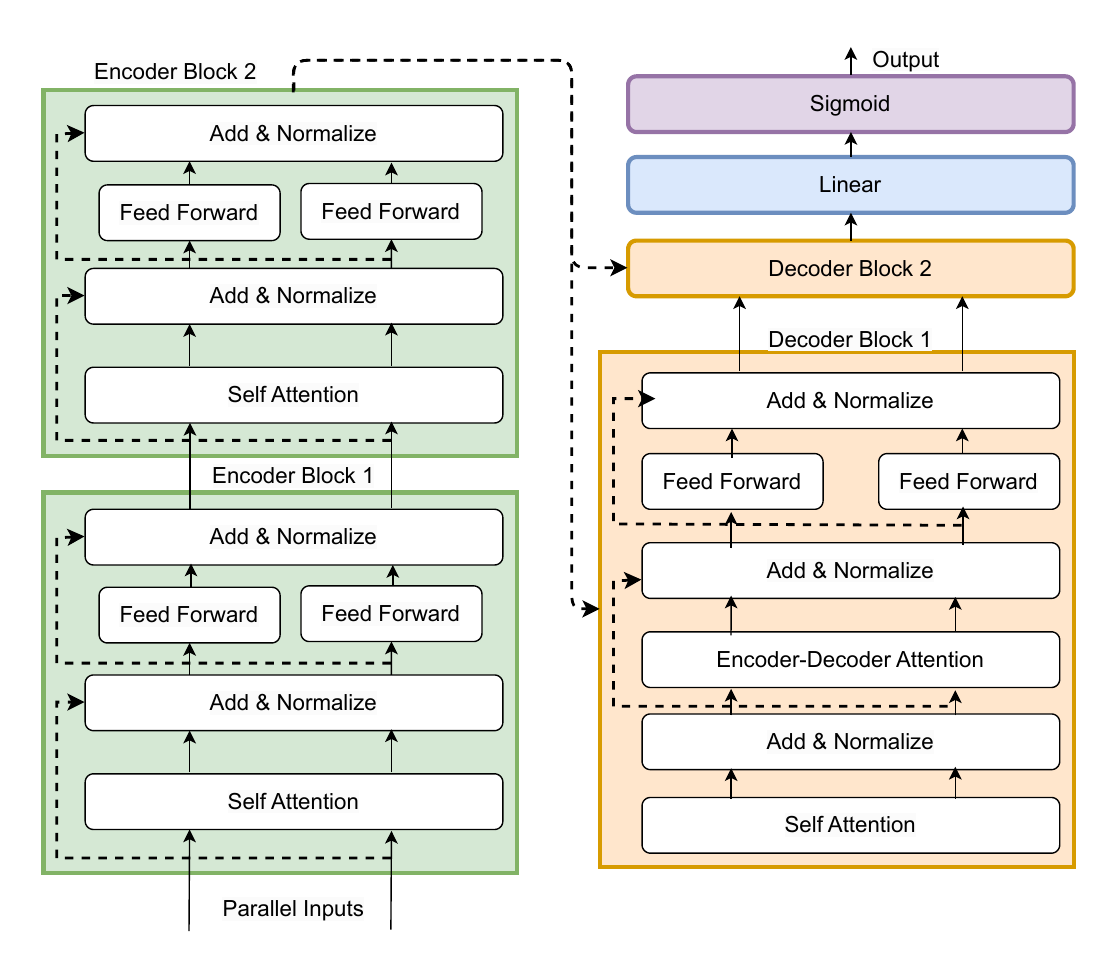}}
\caption{Detailed Structure of the \textit{FTN-TransformerNet}}
\label{fig: Structure of Transformer}
\end{figure}

Fig~\ref{fig: Simplified FTNTransformerNet Structure} shows the model diagram of the proposed \textit{FTN-TransformerNet} equaliser. 
We use Tensorflow to build our Transformer systems. We have an Input Layer of size $L$. 
Fig~\ref{fig: Structure of Transformer} gives the detailed architecture of the Transformer model that we use. 
All the inputs are provided as inputs to the encoder of the Transformer in parallel. 
This is followed by a Self-attention layer. The output of the self-attention block is added to the inputs of the self-attention block. 
This is followed by a set of Feed-forward networks and a ResNet-type architecture where the inputs and outputs are added. 
This completes an Encoder. 
The output of the first encoder is the input to the next Encoder. 
As shown in Fig~\ref{fig: Structure of Transformer}, our model has 2 encoders. 
The output is then sent to the 2 decoders in parallel. 
We cascade the 2 decoders followed by a linear and a sigmoid layer at the end. 
Typically, the Transformer architecture ends with a softmax layer that allocates probabilities across multiple categories. 
However, we simplify the model significantly by opting for a sigmoid activation that allocates a single probability value to an output class. 
This is possible as we have only two possible cases. Namely, 0 and 1 correspond to the symbols -1 and 1. Effectively, the output is the probability of 1. 
We use a low-complexity model of the transformer that has fewer parameters than the LSTM and BiLSTM models discussed earlier. 
This ensures swift inference operation by the equaliser model. This low complexity model has its own drawbacks,. 
Table~\ref{tab: Model Parameters Attn} details the parameters of the \textit{TransformerNet} model.

\begin{table}[]
\centering
\caption{ Model parameters of \textit{FTN-TransformerNet}}
\begin{tabular}{| c|c  |} 
 \hline
 Input shape&$L$\\
\hline
 Attention head size&$N$\\
\hline
 No of Attention heads& 4\\ 
 \hline 
 No of Feed Forward Networks&2\\
\hline
 Feed-forward dimension& $N$\\ 
 \hline
 No of Layers in Feed Forward Network&1\\
 \hline
 No of Transformer Encoder blocks& 2\\
\hline
 No of Decoder blocks& 2\\
\hline
 Dropout& 0.4\\
\hline
\end{tabular}
\label{tab: Model Parameters Attn}
\end{table}

\subsection{Training And Testing The Models}\label{subsec: training-and-testing-the-models}

In this subsection, we detail the methodologies and the parameters to train the models we use in this paper. The models discussed were trained using an Adam optimiser with a standard learning rate of 0.001 unless otherwise stated. 

\begin{figure}
\centerline{\includegraphics[width=0.9\linewidth]{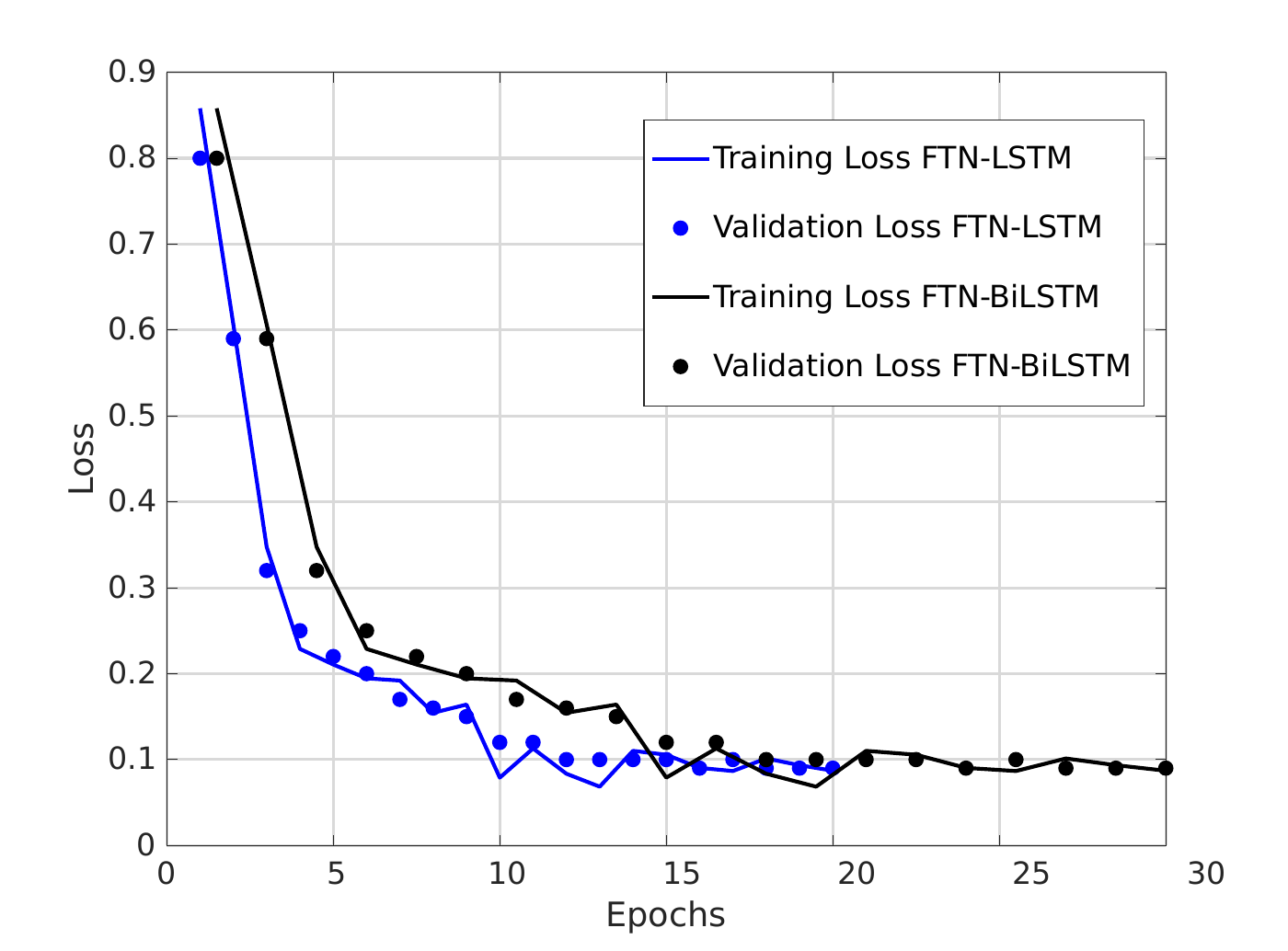}}
\caption{Training and Validation Loss for \textit{FTN-LSTM} and \textit{FTN-BiLSTM}.}
\label{fig: Training and Validation Loss for FTN-LSTM}
\end{figure}

We train the models using a supervised learning approach. 
While training, we set the observation windows ($Y_k $) mentioned in Sec~\ref{sec: The-NN-based-Sliding-window-detection} as the inputs and the corresponding transmitted symbols ($ x[k]$) as the labels. 
Table~\ref{tab: Training and Testing Parameters} details the training and testing setup for both the unidirectional and bidirectional LSTM models. 
To ensure that our datasets are diverse enough, we use a training dataset of 50 to 100 times the number of parameters in the model.
This results in a dataset of size 10 million for both the LSTM and Bi-LSTM models.
The testing dataset is the same size as the training dataset.
This allows for the testing data to be large enough to reliably measure small BER values such as $1.0 \times 10^{-06}$. 
To ensure that our models are robust to noise, we train them over a range of SNRs varying from 6 dB to 12 dB for $\tau = 0.5$ and from 8 dB to 20 dB for $\tau = 0.35$. This is because the impact of ISI is more significant for smaller values of $\tau$  and a higher SNR is necessary to obtain the BERs of interest.  
Fig.~\ref{fig: Training and Validation Loss for FTN-LSTM} shows how the training and validation losses evolve over epochs. 
We find that the losses do not drop significantly after the $17^{th}$ epoch, and thus we employ early stopping after the $20^{th}$ epoch for the unidirectional model. Similarly, for the bidirectional model, it does not drop significantly after the $25^{th}$ epoch, thus we stop after the $30^{th}$ epoch.

\begin{figure}
\centerline{\includegraphics [width = 0.9\linewidth]{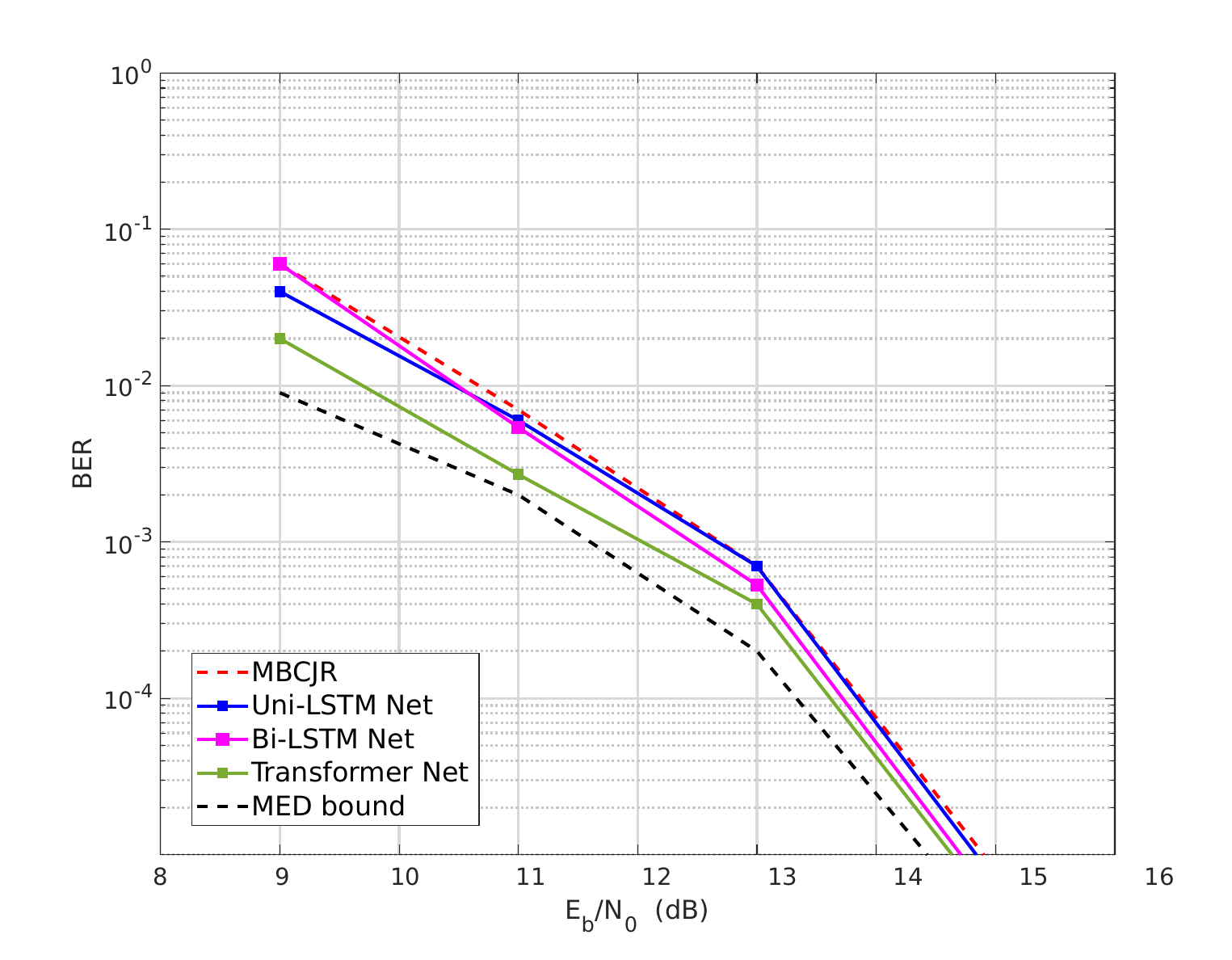}}
\caption{Loss Curves for \textit{FTN-TransformerNet}}
\label{fig: Loss Curves FTNTransformerNet}
\end{figure}

Table~\ref{tab: Training and Testing Parameters} details the training and testing setup of the Transformer model. 
The range of SNRs remains the same as the RNN models. 
Fig.~\ref{fig: Loss Curves FTNTransformerNet} shows the evolution of training and validation losses over epochs. 
Note that the model needs more training data than the Bi-LSTM-based structure. 
Also, the losses decay more slowly than the RNN models. 
This is a relatively simple transformer model, and as such, it does not learn very well in the initial stages of training. 
However, upon enough iterations, the losses eventually start to decrease significantly after 50 epochs. 
And finally, the losses do not drop after the $340^{th}$ epoch and thus we stop after the $400^{th}$ epoch.

\begin{table*}[]
\centering
\caption{ Training and Testing parameters of the models}
\begin{tabular}{| c|c|c  |c|c|}\hline
  &Unidirectional RNNs& Bidirectional RNNs& \multicolumn{2}{|c|}{Transformer}\\\hline 
 \hline
 Training data size  &20,000,000&20,000,000& \multicolumn{2}{|c|}{60,000,000}\\
\hline
 Testing data size     &20,000,000&20,000,000& \multicolumn{2}{|c|}{120,000,000}\\
\hline
 Training max epochs  &300& 300& \multicolumn{2}{|c|}{400}\\
\hline
 Training SNR range  &6-15 dB or 8-20dB & 6-15 dB or 8-20dB & \multicolumn{2}{|c|}{6-15 dB or 8-20dB}\\
 
\hline
 Number of Parameters& 50000& 120,000& \multicolumn{2}{|c|}{400,000}\\\hline
\end{tabular}
\label{tab: Training and Testing Parameters}
\end{table*}

The unidirectional LSTM model is the quickest, followed by the bidirectional LSTM and the transformer takes much longer. The transformer has a larger number of parameters, and so a much larger dataset is required to train it.

\subsection{Training Dataset and Generalisation Concerns}
\label{subsec: Training-Dataset-and-Generalisation-Concerns}

A major concern with any data-driven model is the case of overfitting. 
In other words, the model "rote-learns" the information it has encountered in training but fails to generalise for what it has not been exposed to. 
As such, a critical question arises as to how much of the sample space the model ought to be trained on. 
For large sample spaces, the problem becomes even more difficult. 

Specific to our scenario, even the relatively benign case of $\tau =0.50$ leads to a window size $L' = 28$. 
That implies a total of $2^{28}$  (>$250$ million) unique input combinations. 
This is a scenario where training on the entire input space is prohibitively expensive and this worsens for smaller packing ratios. 
Naturally, a question arises regarding how much data we must have in our training set. 
More detailed analysis is in Sec~\ref{subsec: A-generalisation-concern}.

\section{Complexity Analysis}
\label{sec: Complexity-Analysis}
We compare the computational complexities of the sliding Feed Forward, the RNN detector, the DFE, the Viterbi Decoder (VD), MAP receivers and Transformers for decoding a sequence of length n, considering the parameters sliding channel memory length $(M)$, equaliser taps $(T)$ and the number of states in the trellis $(N = 2^M)$. 
Table~\ref{tab: Receiver_Complexity} lists the complexities of the various receivers.
The DFE’s complexity is $O(nT)$, which depends linearly on the number of taps in the equaliser filter. and the sequence length. 
The sliding RNN detectors have a complexity of $O(R_{U}L'(n - L' + 1))$, that is, the complexity grows linearly with the sequence length $(n)$ and the sliding window size $(L')$. 
Similarly, for the Bi-directional, the complexity is $O(R_{B}2L'(n - L' + 1))$.
For the Transformers, the complexity is $O(R_{T} L'(n - L' + 1))$ and $R$ denotes that though the size of the sliding window is unchanged, 
As $n$ increases, the complexity also grows linearly, but it’s bounded by the window size. 
The VD and MAP receiver complexity is given by $O(nN)$ and $O(2n2^M)$ respectively. 
As channel memory increases, the number of states in the trellis grows exponentially, leading to an exponential increase in complexity for the VD and MAP receivers. 
For the FTN-induced ISI, the effective channel is long and the computational complexity becomes impractical due to its exponential growth. 
The reduced MAP receivers given in \cite{Prlja_turbo} have a lower complexity but it is still high. 
As such, the sliding RNN detectors are more computationally efficient alternatives for decoding FTN signals.
The Transformers also have lower complexity than the traditional receivers but higher than that of the RNN receivers.
The 3rd column in Table~\ref{tab: Receiver_Complexity} gives the asymptotic complexity of the entire system.
For ease of comprehension, it is good to note that $2^M >> 2^{M'} \sim R.L'^3$, where $R>>1$. $R \sim 16$.
We shall discuss the performance of these receivers in Sec~\ref{sec: Results} and analyse if the complexity tradeoffs are justified. 

\begin{table}
    \centering
    \begin{tabular}{|c|c|c|} \hline 
         Model& Algorithm Complexity &Total Asymptotic Complexity\\ \hline 
         DFE& $O(nT)$ &$O(nT)$\\ \hline 
         Viterbi& $O(n2^M)$   &$O(n2^M)$\\ \hline 
         MAP& $O(2n2^M)$  &$O(2n2^M)$\\ \hline 
         Reduced MAP& $O(n2^{M'})$    &$O(n2^{M'})$\\ \hline 
         LSTM& $O(L'(n-L'+1))$&$O(nL'^3)$\\ \hline 
         Bi-LSTM& $O(2L'(n-L'+1))$&$O(2nL'^3)$\\ \hline 
         Transformers& $O(RL'(n-L'+1))$&$O(RnL'^3)$\\ \hline
    \end{tabular}
    \caption{Complexity of Receivers}
    \label{tab: Receiver_Complexity}
\end{table}

\section{Empirical Results}\label{sec: Results}
We consider uncoded BPSK data over an AWGN channel with varying compression factors. 
Further, we pulse shape the transmit signal $x(t) $ with a square root raised cosine pulse having a bandwidth expansion factor $ \beta = 0.30 $. 
In Sec~\ref{subsec: The-Modified-FTN-Channel}, we had discussed the super minimum phase channel and how the approach benefits our receivers. 
In our simulations, we consider the physical channel mentioned above and modify it to obtain its equivalent super minimum phase response as discussed in Sec~\ref{subsec: The-Modified-FTN-Channel}. 

\subsection{The Benchmarks}
\label{subsec: The-benchmarks}
\textbf{Maximum Likelihood Sequence Estimator} (denoted in the paper as \textit{Viterbi} Decoder) is the optimal solution to ISI-based sequence estimation problems. 
Thus, it serves as a valuable benchmark for comparison. 

\textbf{MBCJR Receiver:} The BCJR receiver~\cite{BCJR_OG} designed by Bahl, Cocke, Jelinek and Raviv is the MAP receiver solution to the sequence estimation problem. 
In \cite{Prlja_Reduced}, the authors used the modified channel discussed in Sec~\ref{subsec: The-Modified-FTN-Channel} to reduce the size of the full BCJR receiver and obtained an M-state BCJR solution that yields performance as good as a full BCJR receiver but at reduced complexity. 
As previously discussed,  for small $\tau$ the standard Viterbi or full BCJR are too large to implement. 
Thus, this is the state-of-the-art receiver for small $\tau$ values.
Hence, it is a useful benchmark for us.\\
\textbf{Minimum Euclidean Distance (MED) bounds} obtained in \cite{Prlja_turbo} (denoted as Anderson's bounds, which were in turn obtained by Prlja, Anderson and Rusek from~\cite{Duel-Hallen}) are used as the lower bounds on BER. This serves as the lower bound for all unbiased receivers. \\
\textbf{The Decision Feedback Equalizer}\cite{DFE_OG}: The DFE is an approach which has lower complexity than the above techniques but is prone to error propagation.
In terms of complexity, it is comparable to the RNN models presented in this paper. 
The DFE also works well with minimum phase channels discussed in Sec~\ref{subsec: The-Modified-FTN-Channel}. 
We use the Recursive Least Squares (RLS) and Least Mean Squares (LMS) algorithms to train the \textit{FTN-DFE} equaliser. 
The first block of data uses the RLS algorithm to ensure rapid tap convergence. 
This is used to learn the channel. 
We use the LMS algorithm thereafter to ensure rapid execution. \\

\subsection{BER Performance of Unidirectional LSTMs}
\label{subsec: Relative-BER-Performance}

\begin{figure}
\centerline{\includegraphics [width = 0.9\linewidth]{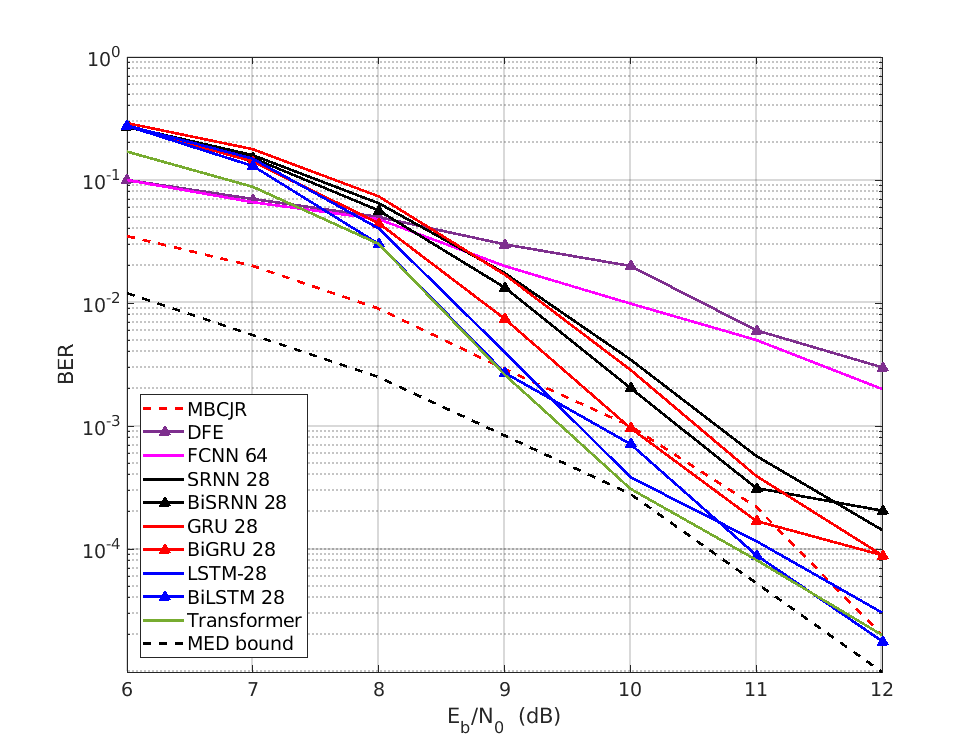}}
\caption{BER comparison of FTN Decoders for $\tau = 0.50$}
\label{BER of FTN Decoders}
\end{figure}

In this case, we evaluate the performance of the FCNN model and the \textit{FTN-LSTM} model.
We benchmark these receivers against the receivers and bounds discussed in the previous subsection. 
Fig.~\ref{BER of FTN Decoders} shows the performance of the receivers. We vary the input size $(L)$ of the \textit{FTN-LSTM.} 
We observe no gains in performance for $L>28$. 
The \textit{FTN-LSTM} emerges as an effective solution. Specifically for $L={28, 46}$ (denoted by \textit{LSTM28} and \textit{LSTM46} respectively), followed by LSTM with $L=16$ (denoted by \textit{LSTM16}). 
The DFE, though computationally less complex, significantly underperforms compared to other models. 
The FCNN model also underperforms the RNN receivers.
The \textit{FTN-LSTM} models are the best solutions, even outperforming Viterbi and MAP receiver performances.
In fact, we observe that the performance of the models is within 0.25 dB of the MED bounds.
Thus, we can conclude that the unidirectional LSTM models achieve nearly optimal performance  $\tau =0.50$ for mid to high SNR regimes.

\subsection{Other Unidirectional RNNs}
\label{subsec: Other Uni-RNNs}
In this subsection, we observe how the performance varies across the different RNN cells. 
We compare three kinds of RNN cells to build the receiver. 
We consider the Simple-RNN, the GRU and the LSTM. 
All three networks have the same number of unfoldings and the first layer of all three models returns the cell states as a sequence. 
We use the same training dataset to train all three models. We simulate the models on the same testing dataset. 
The \textit{FTN-LSTM} outperforms the \textit{FTN-GRU} and it outperforms the \textit{FTN-SRNN}. 
This is in line with expectations as LSTMs can capture more complex relations. 
Note that for a given size of the network, LSTMs have the highest number of parameters among all the three variants of the RNN. 
The LSTM has an extra output: the hidden state. This allows more control over the flow of contextual data, which implies that the model can select how much of the past samples it can forget. 
This allows the \textit{FTN-LSTM} model to learn better than the other models.

\subsection{Performance for different packing ratios}
\label{subsec: Performance-for-different-packing-ratios}
\begin{figure}
    \centering
    \includegraphics[width=0.9\linewidth]{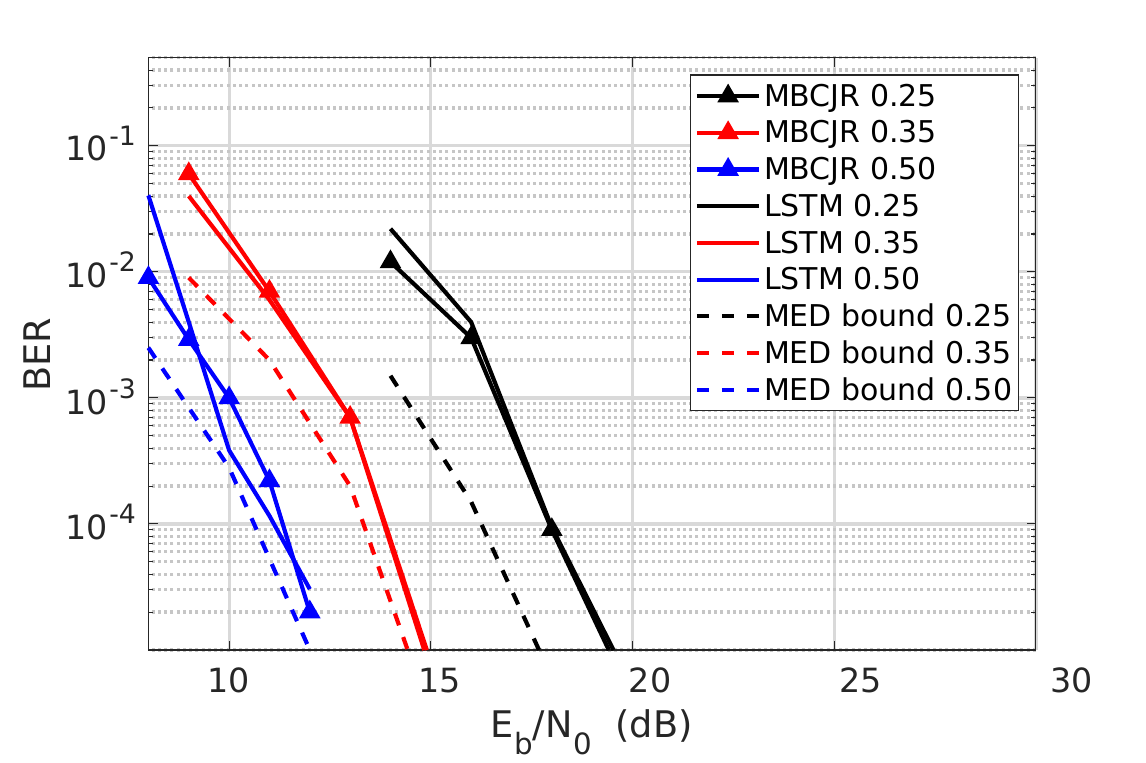}
    \caption{BER Performance across varying packing ratio models}
    \label{fig-varying-packing-ratios}
\end{figure}

In this part, we observe the performance of our RNN-based equalisers for different packing ratios. 
Specifically, we consider the cases of $\tau = 0.5, 0.35$ and $0.25$. 
We train separate models for the different packing ratios. Fig~\ref{fig-varying-packing-ratios} shows the BER performance of the various packing ratios. 
As expected, the RNN equaliser performs best for $\tau = 0.50$ and worsens as $\tau$ reduces to $0.25$.
As $\tau$ reduces, the ISI increases and as a result, the errors increase for a given SNR level.
It is worth noting that for $\tau = 0.35 \text{ and } 0.25$, our models show similar performance as those obtained by the MAP receivers in \cite{Prlja_Reduced}. 

\begin{figure}
\centerline{\includegraphics [width = 0.9\linewidth]{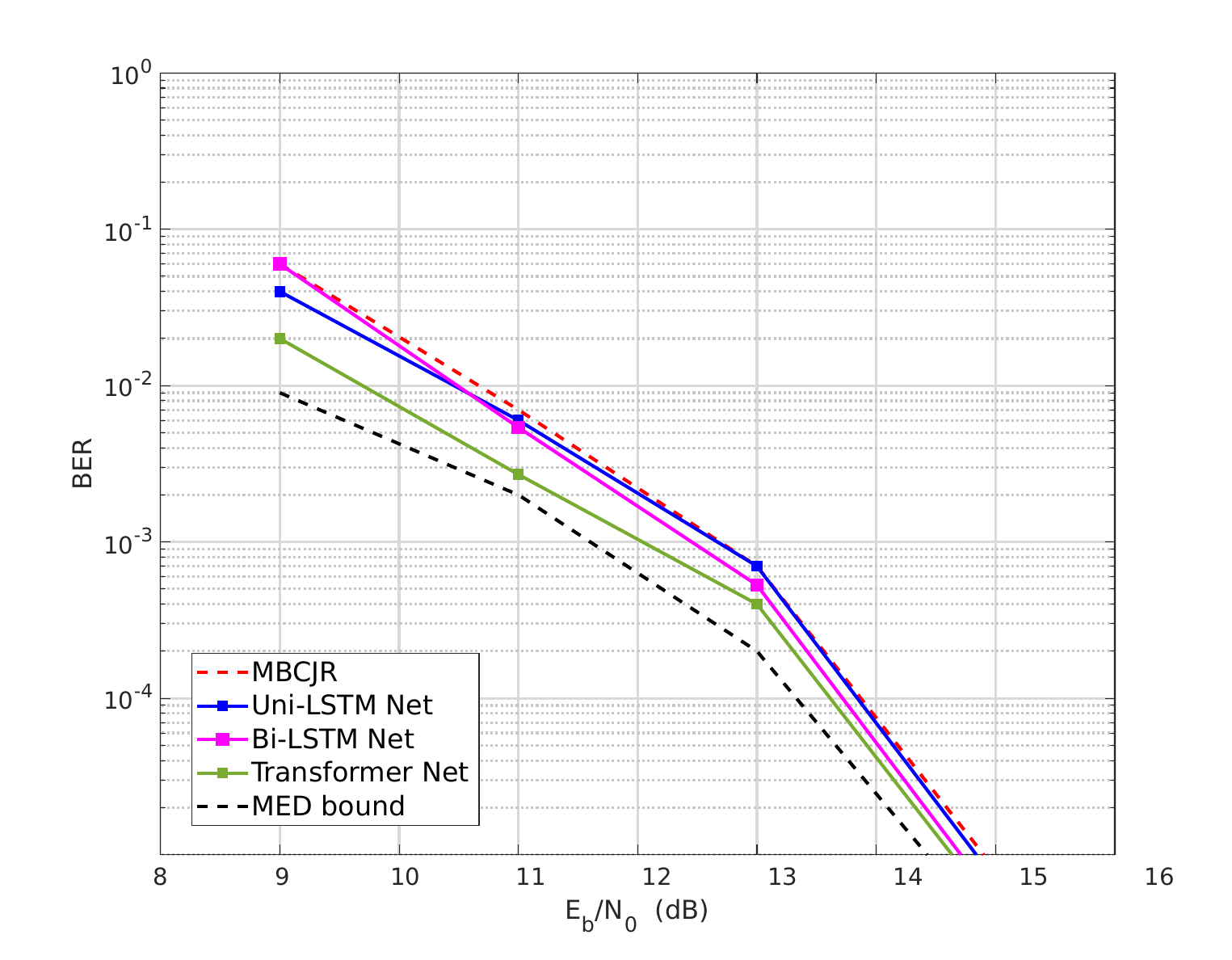}}
\caption{BER comparison of FTN Decoders for $\tau = 0.35$}
\label{fig: BER of FTN Decoders 0.35}
\end{figure}

In this subsection, we study the BER performance of the bidirectional-LSTM models. 
Fig~\ref{BER of FTN Decoders} and Fig~\ref{fig: BER of FTN Decoders 0.35} give the BER plots for FTN packing ratios $\tau = 0.5 \text{ and } 0.35$ respectively. 

We observe that \textit{FTN-BiLSTMNet} (designed using bidirectional LSTM) does not lead to any meaningful gain in BER performance for $\tau =0.5$ but it shows some gain for $\tau = 0.35$. 
It should also be noted that at lower SNRs, it underperforms the uni-directional LSTM. 
The performance may be explained in the context of the modified channel response. 
As depicted in Fig.~\ref{fig: Prlja_Channels}, for $\tau =0.5$, the energy of the channel taps rises very quickly and most of the energy is contained within the first 3 major taps. 
So, there is very little energy to be obtained by the feedback system offered by the Bi-LSTM-Net. 
However, for $\tau = 0.35$, the energy rises less aggressively and there is significant energy to be obtained by the feedback system of the Bi-LSTM-Net. 
Thus the BiLSTMNet can use the predictions on the future symbols to retrospectively alter predictions on past symbols. 
This leads to improved performance by the Bi-LSTM-Net.

\begin{figure}
\centerline{\includegraphics [width = 0.9\linewidth]{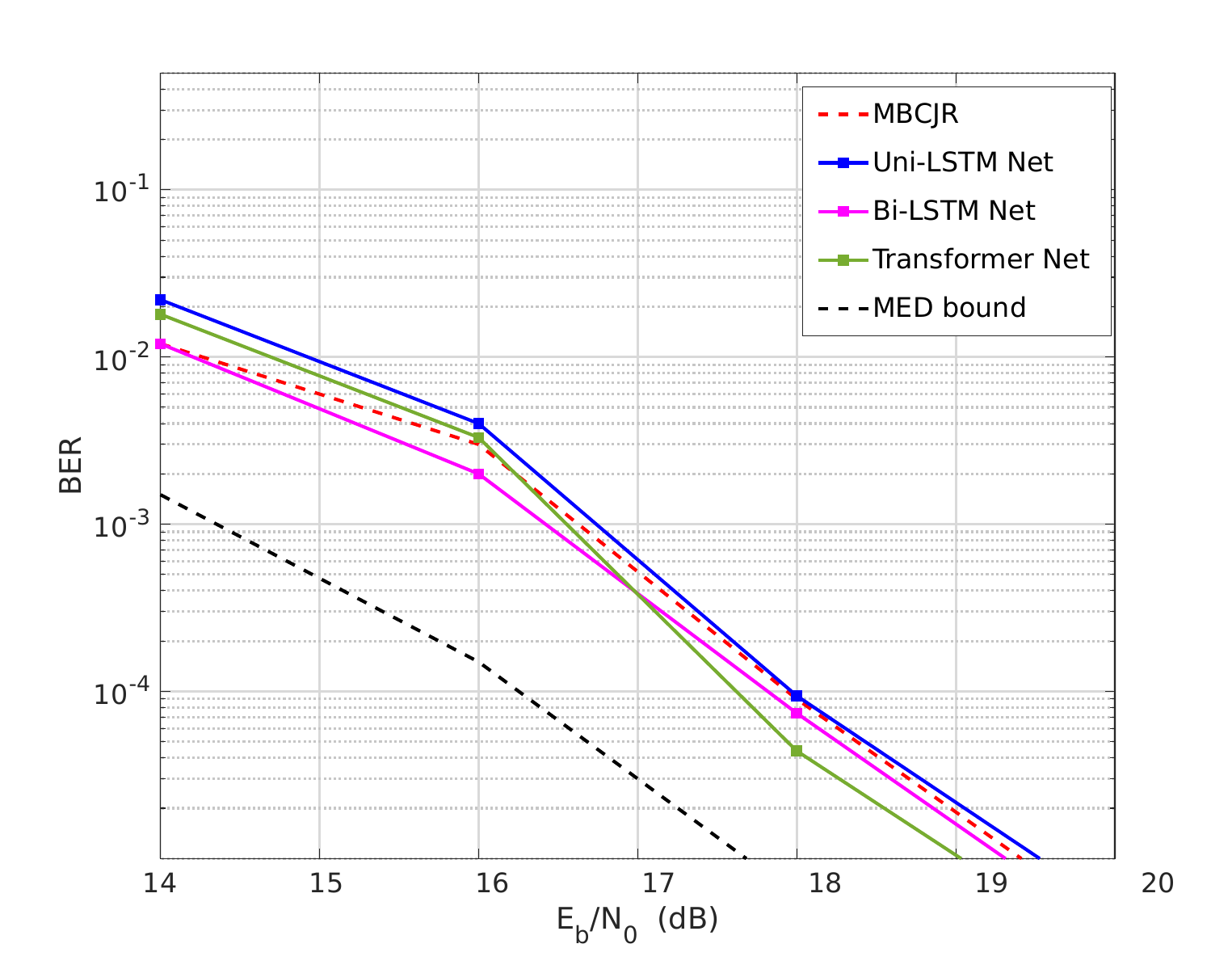}}
\caption{BER comparison of FTN Decoders for $\tau = 0.25$}
\label{fig: BER of FTN Decoders 0.25}
\end{figure}

\begin{figure}
\centerline{\includegraphics [width = 0.9\linewidth]{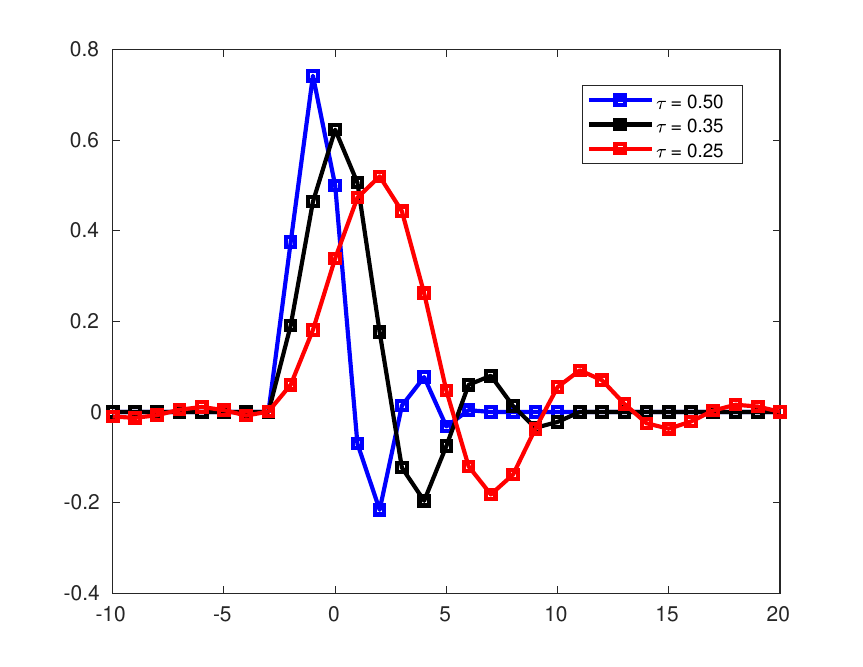}}
\caption{Modified Channels}
\label{fig: Prlja_Channels}
\end{figure}

\subsection{Transformer Network}
\label{subsec: Transformer Network}
In this subsection we discuss the performance of the \textit{FTN-TransformerNet} detailed in Sec.~\ref{subsec: Transformer Network}.  
There is a marginal gain in BER performance for $\tau =0.5$ as shown in Fig.~\ref{BER of FTN Decoders}. 
Especially at lower SNRs. 
This can be attributed to the fact that transformers operate on the principle of attention. 
Thus, if a symbol is strongly affected by noise, the model can use information from other parts of the observation window that may be less noisy. 
As the SNR improves, this effect becomes less pronounced.
For $\tau = 0.35$ and $\tau=0.25$, \textit{FTN-Transformer-Net} shows a significant improvement. 
Again, this is more pronounced at lower SNRs. 
As mentioned previously, for $\tau = 0.35$, there is significant energy in the feedback path. 
It allows the attention models' parallel processing mechanism to use "future" symbols to make better predictions about present and past symbols. 
So for $\tau = 0.35$ and $\tau=0.25$, the \textit{Transformer-Net} performance is closer to the bound as demonstrated in Fig.~\ref{fig: BER of FTN Decoders 0.35} and Fig~\ref{fig: BER of FTN Decoders 0.25}.
However, this performance improvement comes at a cost of higher complexity, a larger model, longer time to train and potentially increased costs. 
We leave it up to the designer to decide if the costs of such a model are outweighed by the benefits of the higher performance.

\subsection{Sensitivity to Training Dataset SNR}
\label{subsec: Sensitivity-to-Training-Dataset-SNR}
In this subsection, we shall explore the impact of the training dataset SNR on the models.
Any neural network model is sensitive to the training dataset and thus to the SNR of the observed data in the training dataset.
We show experiments for $\tau = 0.5$ with the LSTM model, but similar results can be obtained for other models too.

We train the models using a supervised learning approach. 
During training, we set the observation windows ($Y_k $) as the inputs and the corresponding transmitted symbols ($ x[k]$) as the labels. 
Table~\ref{tab: Training and Testing Parameters} details the training and testing setup for the \textit{FTN-LSTM} model. 
To test our model's resilience to noise, we train multiple instances of the model each for a different value of SNR. The SNRs vary from 6 dB to 12 dB. 

\begin{figure}
    \centering
    \includegraphics[width=0.9\linewidth]{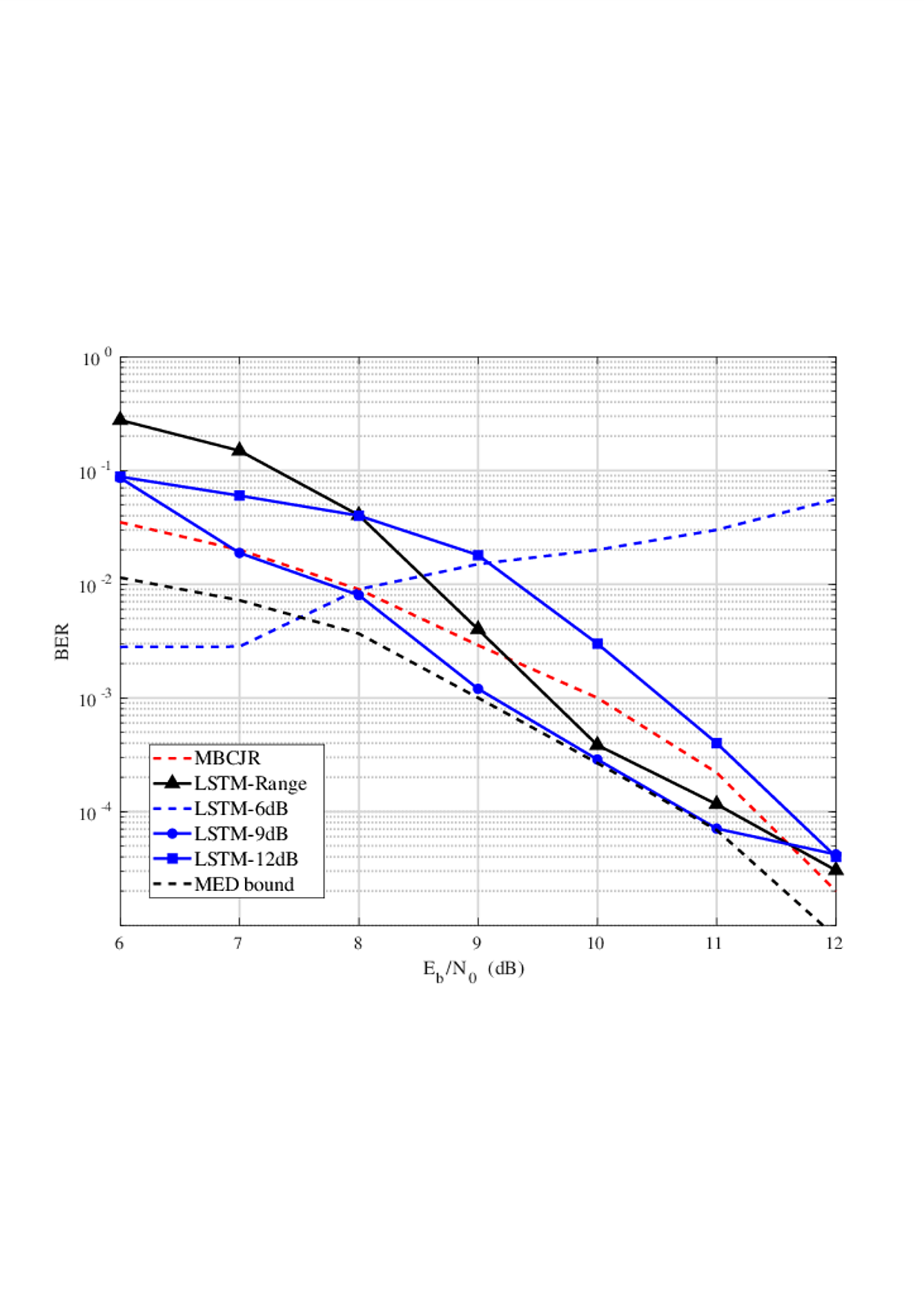}
    \caption{BER Performance of SNR specific models}
    \label{fig: Training SNR impact on BER}
\end{figure}

\textbf{Case 1: SNR specific models: }Fig~\ref{fig: Training SNR impact on BER} shows how a model trained for a specific SNR value performs over a range of SNRs in the testing dataset. 
As we can see from the figure, the model trained with 9dB SNR in training data performs the best. 
The model trained with 6dB SNR encounters an excessively noisy dataset. 
Even though it performs well in the low SNR regime, it fails when the SNR is high. 
Thus, it does not learn the underlying signals very well. 
However, the model trained for 12dB, performs poorly at low SNRs but performs well for high SNRs. 
This model can be understood to be very sensitive to the data it encounters. 
On the other hand, the model trained at 9dB SNR is at a "sweet spot". 
This model delivers good performance in both mid and high-SNR scenarios. 
It somewhat underperforms the 6dB model in low SNR cases but continues to provide performance comparable to the 12dB model even at high SNRs.

\textbf{Case 2: Training over a range of SNRs: }
Fig~\ref{fig: Training SNR impact on BER}  also shows (black solid line) how a model trained over a range of SNR values performs over a range of SNRs in the testing dataset. 
We train the model with training data with SNRs varying from 6 dB to 12 dB. 
As we can see from the figure, the model trained across the range underperforms the models in the previous case in almost all cases. 
However, the model continues to be robust across the range of SNRs in the testing dataset.

Though our analysis is for 0.5 only, a similar approach could be used to train for other values of $\tau$. The caveat being that the SNR range needs to be appropriately altered.

\subsection{Sensitivity to Sampling Offset}
\label{subsec: Sensitivity-to-Sampling-Offset}

\begin{figure}
\centerline{\includegraphics[width=0.9\linewidth]{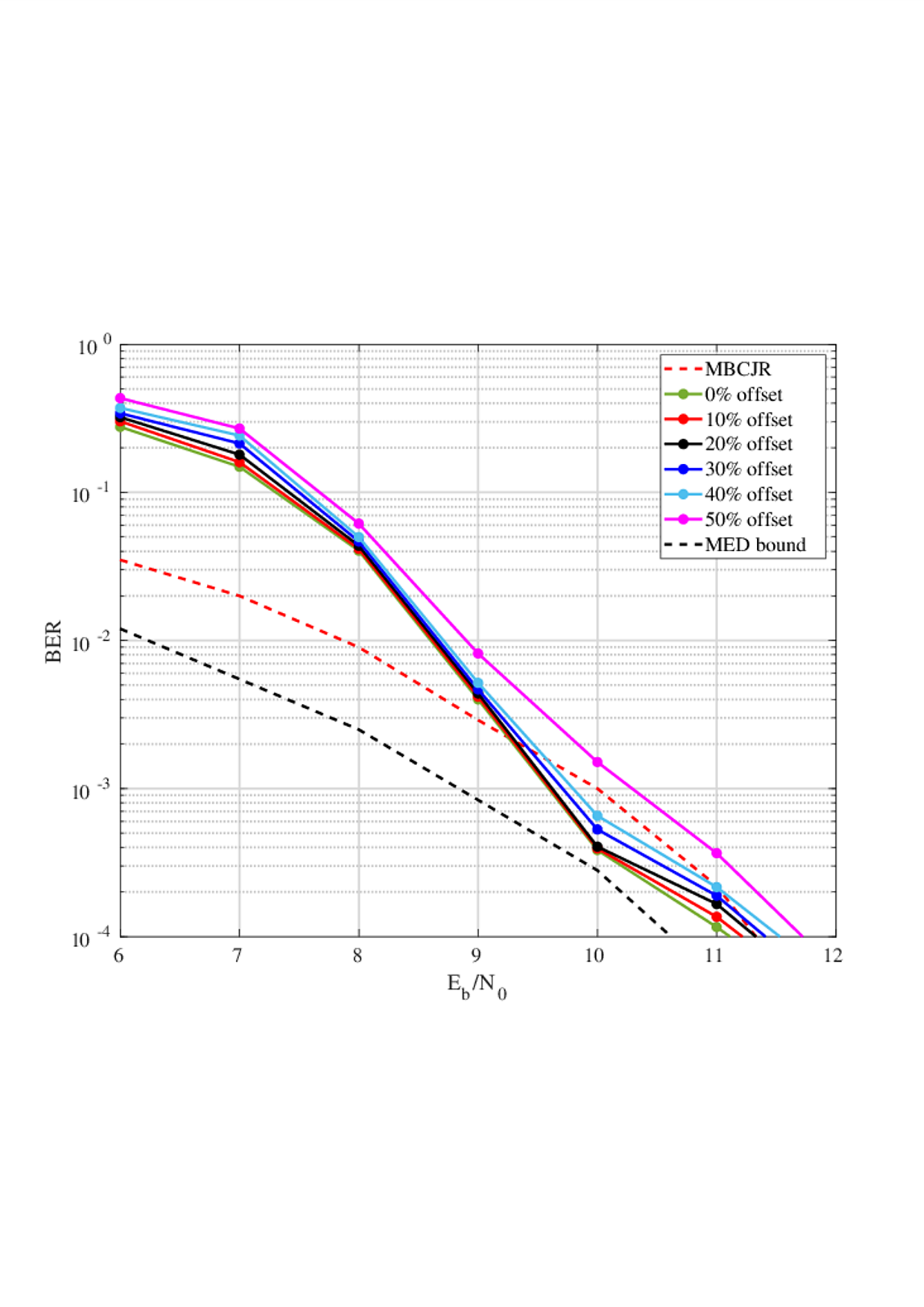}}
\caption{BER performance with sampling offset for $\tau = 0.50$ with LSTM equaliser}
\label{BER performance with Sampling Offset LSTM}
\end{figure}

In the previous experiments and in the literature\cite{Prlja_Reduced,Prlja_turbo,Anderson_turbo,Self_1}, it was assumed that the signals were sampled at the ideal sampling points. 
In this subsection, we investigate the impact of sampling offset on the BER performance of our models. 
We introduce synchronisation offsets into the system. 
The models are trained with a clean training dataset without any offset but offsets are introduced while inference.
We consider the Long Short-Term Memory (LSTM) cell and quantify the impact of offset on BER performance.
Fig.~\ref{BER performance with Sampling Offset LSTM} shows the BER performance over a range of offsets. 
Our LSTM models are robust to offsets of up to $40\%$ of the symbol duration.

The models are resilient against non-linearities in sampling. In a real-life scenario where the waveform is sampled, minor errors in sampling will not impact the performance.

\subsection{A Generalisation Concern}
\label{subsec: A-generalisation-concern}
As discussed in Sec~\ref{subsec: Training-Dataset-and-Generalisation-Concerns}, a concern arose about the ability of our models to generalise well across sequences that it did not train on and we investigate this now.\\
\textbf{Case 1: Run Length Constraints} 

\begin{figure}
\centerline{\includegraphics[width=0.9\linewidth]{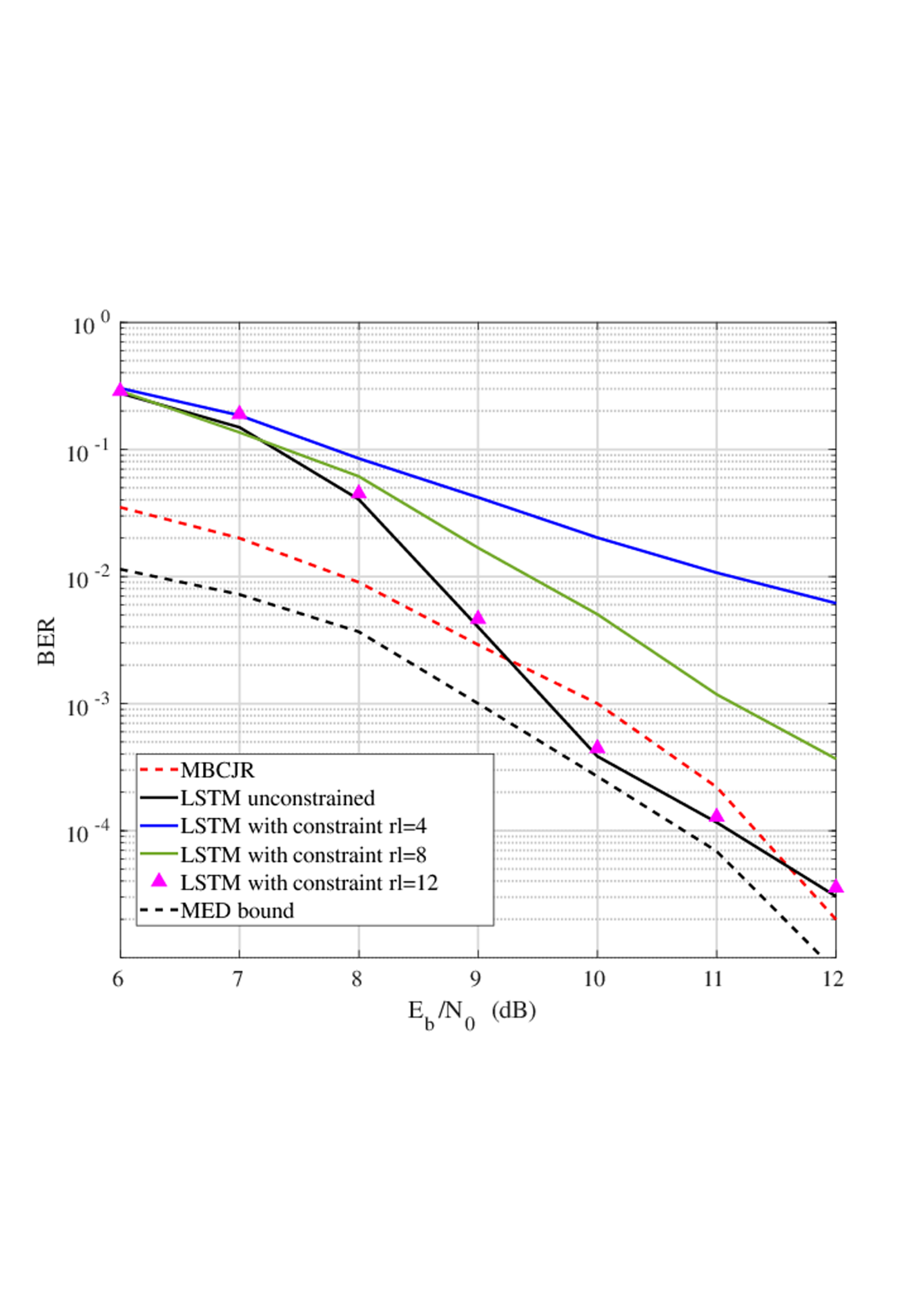}}
\caption{BER performance with Biased Training for $\tau = 0.50$}
\label{fig-BER performance with Biased Training}
\end{figure}

We enforce run-length constraints on the training dataset to ensure that certain sequences are absent in the training dataset. 
We observe the performance of the model trained on this constrained data. 
The testing dataset is unconstrained and has some sequences that the model was not trained on. Fig~\ref{fig-BER performance with Biased Training} shows how enforcing constraints on training reduces performance. 
We have used Simple RNN Cells for this experiment. $RL12$ implies a run-length constraint of 12. 
That is, a sequence of more than 12 consecutive $+1$s or $-1$s will not occur in the training dataset. 
Similarly, $RL8$ and $RL4$ are stronger constraints that prohibit sequences of $8$ and $4$ consecutive symbols respectively. 
As we observe, the models that saw less variation in training data, perform worse. 
This is why $RL4$ performs the worst followed by $RL8$ and $RL12$. And $RL12$ performs almost as well as the default.
This indicates that our models do not perform well if the dataset is inherently biased.
This further implies that our unconstrained dataset has ample variations in the sequences.

\textbf{Case 2: Completely Dissimilar Training and Testing Datasets} 

\begin{figure}
\centerline{\includegraphics[width=0.9\linewidth]{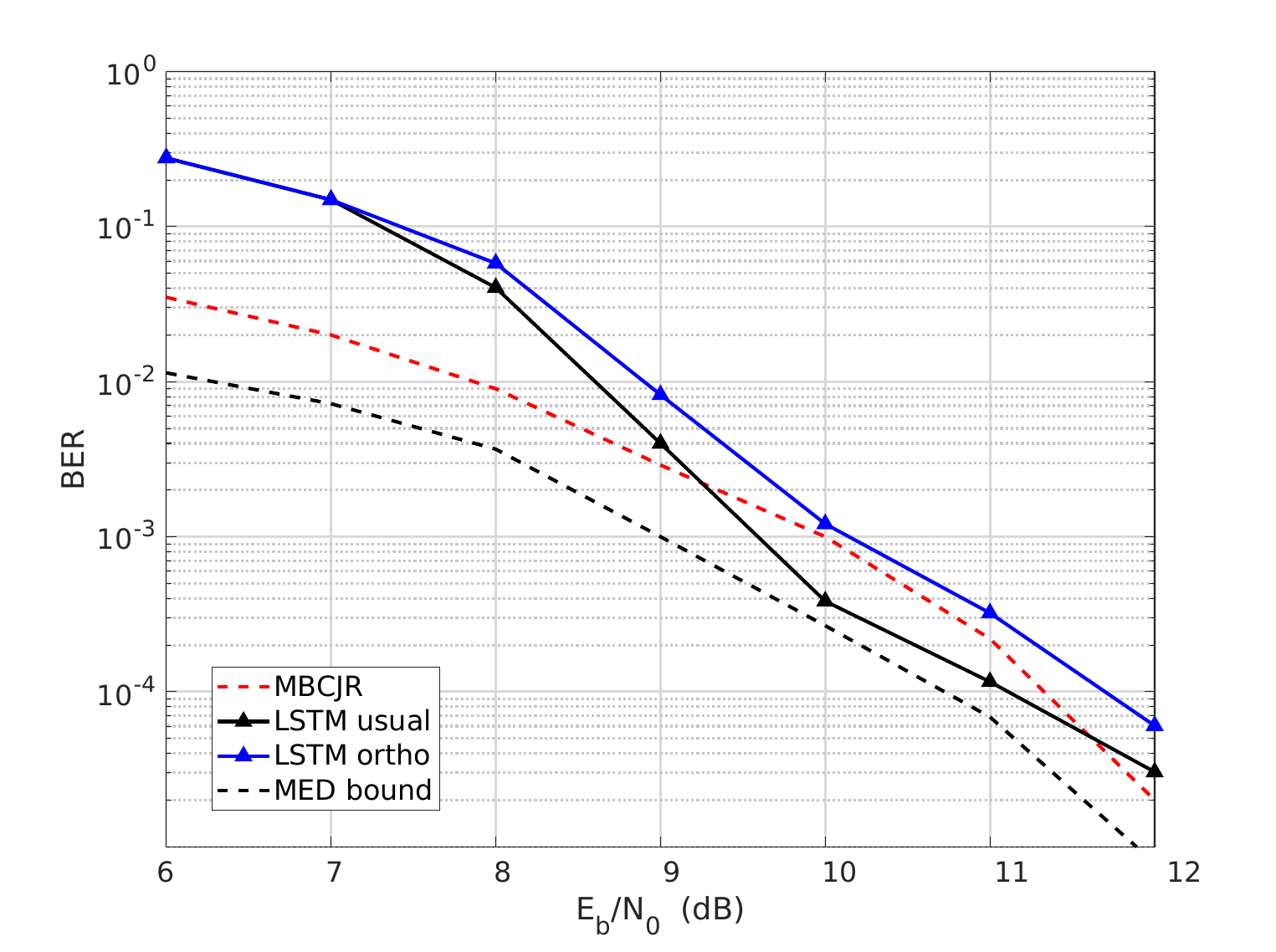}}
\caption{BER performance with Orthogonal Training and Testing Datasets for $\tau = 0.50$}
\label{fig-BER performance with Orthogonal Training and Testing Datasets}
\end{figure}

In this experiment, we do not constrain the training dataset. 
But we pick and choose the sequences in the testing set. 
We ensure that the sequences in the testing dataset are completely new to the sequences in the training dataset. Fig~\ref{fig-BER performance with Orthogonal Training and Testing Datasets} captures the BER performances of the model. 
As we can observe, there is some loss in performance. 
However, we cannot certainly deduce if this implies a lack of generalization. 
Because constructing a testing dataset orthogonal to the training dataset leads to non-contiguous sequences. 
This is a major hindrance as the model now performs a block detection instead of a sliding window detection. 
In this light, it is difficult to comment if the loss observed is due to non-contiguity or lack of generalization.
\\\textbf{Case 3: Varying Training Dataset Sizes} 

\begin{figure}
\centerline{\includegraphics[width=0.9\linewidth]{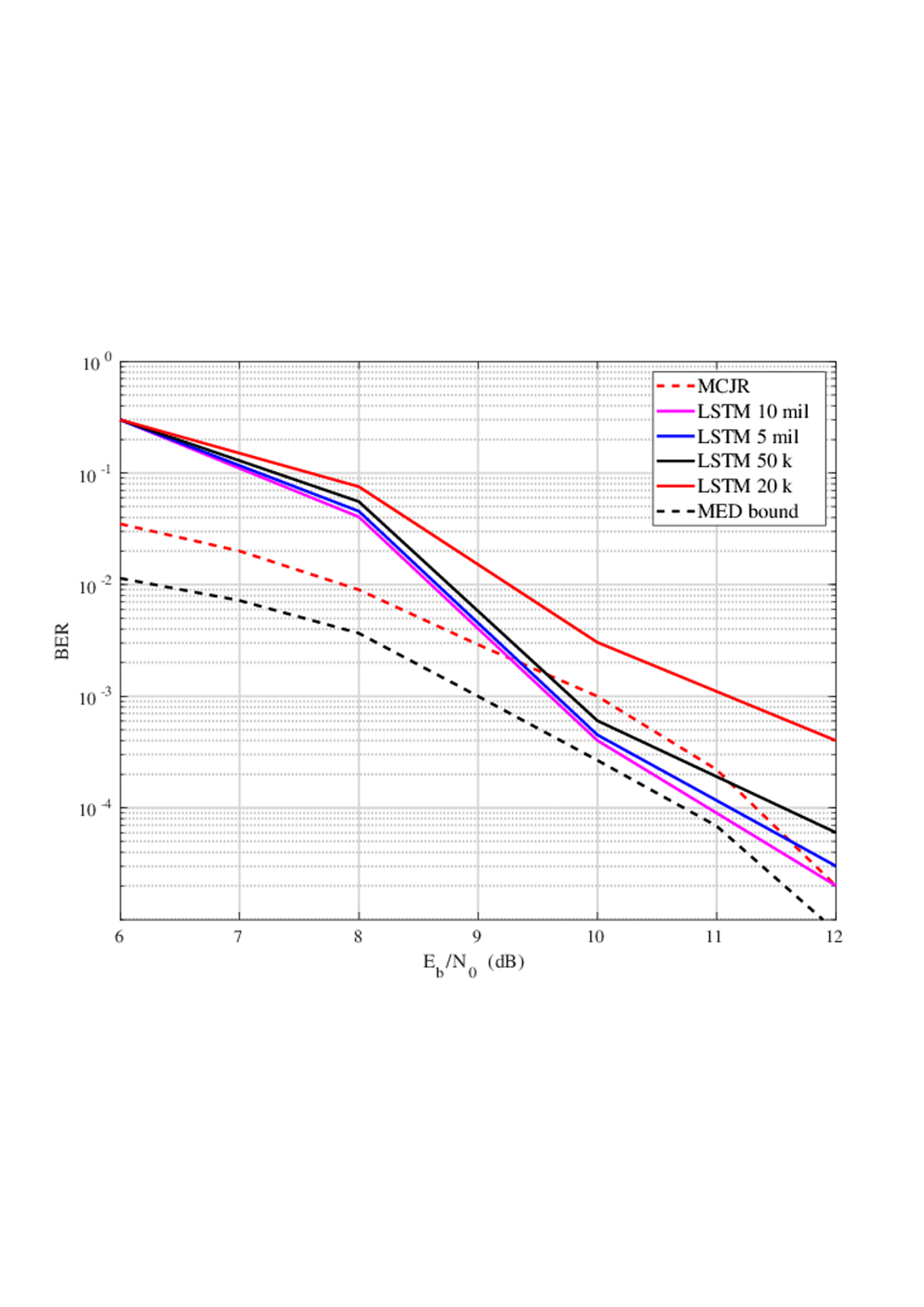}}
\caption{BER performance with Varying Training Dataset size for $\tau = 0.50$}
\label{fig-BER performance with Varying Training Dataset size}
\end{figure}

In this experiment, we vary the training dataset size. We use 4 different sizes of training datasets. 
The sizes vary from 10 million to 20000 unique sequences. 
Fig~\ref{fig-BER performance with Varying Training Dataset size} shows us the relative BER performances. 
As we can observe, there is little loss in performance with the reduced size of the training dataset up to 50000 sequences.
But what is essential is that we have enough noise realisations.
Nevertheless, there is a significant loss in performance for 20000 sequences.
That is, the dataset is too small.
\subsection{Higher Modulation Schemes}

\begin{figure}
\centerline{\includegraphics[width=0.9\linewidth]{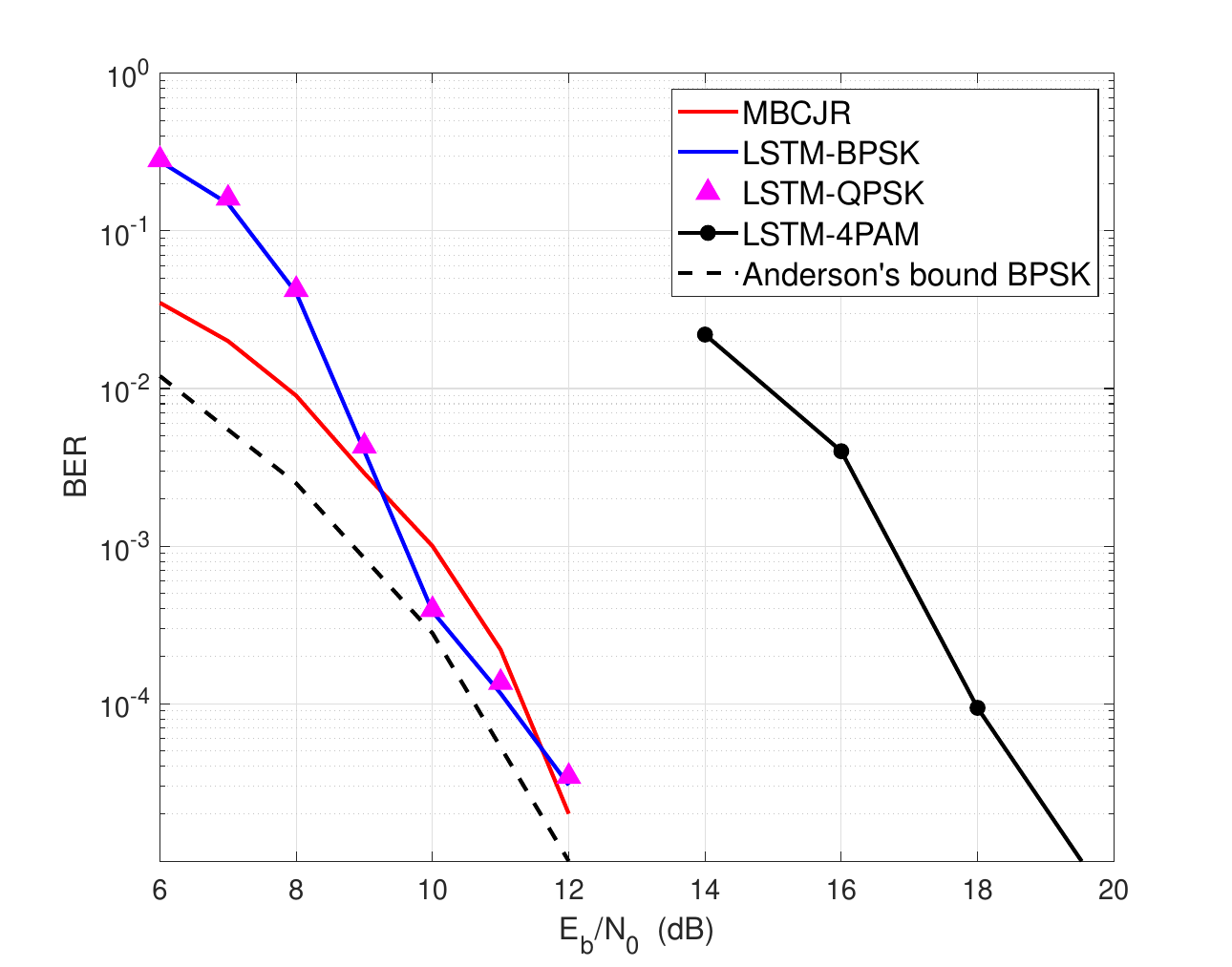}}
\caption{BER performance with Higher modulation for $\tau = 0.50$}
\label{fig-BER performance with Higher modulation}
\end{figure}
We examine the implications of higher modulation schemes.
We observe from Fig~\ref{fig-BER performance with Higher modulation} that for QPSK performance is similar to BPSK.
This is expected as the QPSK scheme is essentially two parallel BPSK schemes that are orthogonal to each other.
Specifically, we simulate QPSK in the NN models as two sets of disconnected inputs to the models with every layer size scaled up 2 times.
Thus the second set of symbols have no impact on the performance on the first and vice versa.
On the other hand, for the coherent 4-PAM, the performance is significantly worse.
This is because the possible sources of interference is significantly higher.
Also, the complexity of the network is higher.
For the 4PAM, we need to replace the binary cross-entropy with categorical cross-entropy.
Also, a 4-PAM with $\tau = 0.5$ transmits 4 bits every $T_0$ secs.
This is equivalent to a BPSK with $\tau = 0.25$.
The performances are also similar.
Thus it is better to use a QPSK than a 4PAM.
Though the first comes at the cost of twice the number of channels.
\section{CONCLUSION}
\label{sec: Conclusion}
The focus of this work is the equalisation of heavily compressed Faster than Nyquist (FTN) signals using non-linear equalisers. 
We proposed multiple DNN methods, namely a)Recurrent Neural Networks (RNNs) and b) Transformers. We assess the efficacy of these models in achieving low-complexity, low-latency, symbol-by-symbol detection of FTN signals with significant reliability. 
Additionally, we have analysed the impact of varying RNN cell types, RNN depths, and block sizes on performance. 
Our investigation has encompassed the assessment of unidirectional RNNs in the presence of non-idealities such as sampling offset, highlighting their robustness under varying Signal-to-Noise Ratio (SNR) conditions. 
Furthermore, we have evaluated the generalisability of the models and extended their applicability to multiple FTN packing factors. 
Our exploration has also delved into the use of explicit feedback from past decisions through bidirectional models, demonstrating instances where they outperform their unidirectional counterparts. 
Lastly, we have examined the performance of Transformers in simplifying the detection of FTN signals.
We observe that Transformers almost always outperform Bi-LSTMs  which in turn are better than Unidirectional LSTMs. 
And the Unidirectional LSTMs are themselves robust and exhibit more reliable performance than Trellis decoders for the detection of highly compressed FTN signals.

\bibliographystyle{IEEEtran}
\bibliography{Bibliography.bib}

\end{document}